\documentclass[aps,pre,twocolumn,superscriptaddress,floats,floatfix]{revtex4}
\usepackage{amssymb,amsmath,amsthm}
\usepackage{graphicx}
\usepackage{color}
\usepackage{subfigure}
\usepackage{epsfig}
\usepackage{rotating}
\usepackage{mwe}
\usepackage{float}
\usepackage{dcolumn,bm}
\usepackage{verbatim}
\usepackage{hyperref}

\begin{document}
	
\title{The Statistical Properties of Superfluid Turbulence in $^4$He from the
Hall-Vinen-Bekharevich-Khalatnikov Model}
\author{Akhilesh Kumar Verma}
\email{akhilesh@iisc.ac.in}
\affiliation{Centre for Condensed Matter Theory, Department of Physics, Indian Institute of Science, Bangalore 560012, India.}
\author{Sanjay Shukla}
\email{ssanjay@iisc.ac.in}
\affiliation{Centre for Condensed Matter Theory, Department of Physics,
Indian Institute of Science, Bangalore 560012, India.}
\author{Vishwanath Shukla}
\email{research.vishwanath@gmail.com}
\affiliation{ Department of Physics, Indian Institute of Technology, Kharagpur, Kharagpur-721302, India.}
\author{Abhik Basu}
\email{abhik.basu@saha.sc.in}
\affiliation{Theory Division, Saha Institute of Nuclear
Physics, Calcutta 700064, India}
\author{Rahul Pandit}
\email{rahul@iisc.ac.in}
\altaffiliation[\\ also at~]{Jawaharlal Nehru Centre For Advanced
Scientific Research, Jakkur, Bangalore, India.}
\affiliation{Centre for Condensed Matter Theory, Department of Physics,
Indian Institute of Science, Bangalore 560012, India.}
\date{\today}
\begin{abstract}

We obtain the von K\'arm\'an-Howarth relation for the stochastically forced
three-dimensional Hall-Vinen-Bekharvich-Khalatnikov (3D HVBK) model of
superfluid turbulence in Helium ($^4$He) by using the generating-functional approach.  
We combine direct numerical simulations (DNSs) and analyitcal studies to
show that, in the statistically steady state of homogeneous and
isotropic superfluid turbulence, in the 3D HVBK model, the
probability distribution function (PDF) $P(\gamma)$, of the ratio $\gamma$  of
the magnitude of the normal fluid velocity and superfluid velocity,
has power-law tails that scale as $P(\gamma) \sim \gamma^3$,
for $\gamma \ll 1$, and $P(\gamma) \sim \gamma^{-3}$, for $\gamma \gg 1$.
Furthermore, we show that the PDF $P(\theta)$, of the angle $\theta$
between the normal-fluid velocity  and superfluid velocity exhibits the
following power-law behaviors: $P(\theta)\sim \theta$ for $\theta \ll
\theta_*$ and $P(\theta)\sim \theta^{-4}$ for $\theta_* \ll \theta \ll
1$, where $\theta_*$ is a crossover angle that we estimate. From our DNSs 
we obtain energy, energy-flux, and mutual-friction-transfer spectra, and
the longitudinal-structure-function exponents for the normal
fluid and the superfluid, as a function of the temperature $T$,
by using the experimentally determined mutual-friction coefficients
for superfluid Helium $^4$He, so our results are of direct relevance 
to superfluid turbulence in this system.

\end{abstract}

\maketitle
\section{ INTRODUCTION}

Over the past three decades, there has been considerable progress in the
characterization of the statistical properties of turbulent fluids by combining
methods from nonequilibrium statistical mechanics and fluid
dynamics~\cite{frisch1995turbulence,RPPramanareview,BoffettaMultiscaling,boffetta2012review}.
By comparison, the study of the statistical properties of turbulent superfluids
is in its infancy; but this field has experienced a renaissance because of
advances in
experiments~\cite{donnelly1998omfdata,maurer1998local,lathrop2011review,bewley2006superfluid,Henn2009bec1st,roche2007vortdensityspectra,guo2010visualization,revvisualqt,salort2011investigation},
and developments in theoretical and numerical
investigations~\cite{berloff2014modeling,l2006energy,nemirovskii2013quantum,shukla2015homogeneous,kobayashi2005kolmogorov}.
The most common experimental system is liquid Helium $^4$He in its superfluid state,
for temperature $T \leq T_\lambda$, the superfluid transition temperature; in
addition, turbulence in superfluid $^3$He and Bose-Einstein condensates (BECs)
is also being
explored~\cite{procacciashell3He,bradley2006decay,parker2005emergence,kobayashi2007quantum}. 

The following models have been employed to study superfluid turbulence: (A) At
the kinetic-theory level there is the model of Zaremba, Nikuni, and
Griffin~\cite{zaremba1999dynamics}.  (B) For weakly interacting Bose
superfluids, we can use a Gross-Pitaevskii description, which is applicable
down to length scales that are comparable to the core size of a quantum
vortex~\cite{nore1997kolmogorov,vmrnjp13,giorgio2011longPRE}. (C)
Vortex-filament models, which are useful at length scales of the order of the
typical separation between quantum
vortices~\cite{schwarz1985three,schwarz1988three,hanninen2014vortex}. (D) the
Hall-Vinen-Bekharevich-Khalatnikov (HVBK) two-fluid model, with
interpenetrating superfluid (s) and normal-fluid (n) components, which
generalizes the two-fluid models of Landau and
Tisza~\cite{landau1941theory,tisza1947theory}, by including a
\textit{mutual-friction} term; the HVBK model provides a good starting point
for the study of superfluid turbulence at length scales larger than several
inter-vortex-separation lengths~\cite{barenghi1992Coutte,roche2009HVBKdns} and
if there is a high density of quantum vortices that align in some regions to
yield a classical vorticity field; however, in experimental flows with quantum turbulence, phenomena such as vortex reconnections~\cite{vinen2002quantum}, which occur at scales comparable to the inter-vortex-separation length, cannot be taken into account by the HVBK model. Measurements on liquid $^4$He have been used
to determine the temperature dependence of the mutual-friction
coefficients~\cite{donnelly1998omfdata}. (E) Wave-turbulence models of superfluid
turbulence~\cite{wtpnasrev,kozik2004kelvin,l2010spectrum} have been used,
\textit{inter alia}, to study Kelvin waves in a turbulent superfluid.

The HVBK description of superfluid turbulence has been successful in obtaining
energy spectra in statistically steady superfluid turbulence, in both three
dimensions (3D) and two dimensions (2D), and in examining the
mutual-friction-induced alignment of superfluid and normal-fluid
velocities~\cite{salort2011mesoscaledns,salort2012energy,shukla2015homogeneous}.
The multiscaling of velocity structure functions and other measures of
intermittency are now being examined both
experimentally~\cite{salort2011investigation,rusaouen2017intermittency,varga2018intermittency},
numerically, and
theoretically~\cite{procacciaintermittencyshellmodel,shukla2016multiscaling,biferale2018turbulent}.
Most theoretical and numerical work on such multiscaling has been restricted to
HVBK-shell-model studies. Furthermore, a precise generalization of the
von-K\'arm\'an-Howarth relations, which have been obtained for classical-fluid
and magnetohydodynamics (MHD)
turbulence~\cite{polyakov1995turbulence,yakhot2001mean,basu2014structure}, does
not seem to be available for superfluid turbulence, to the best of our
knowledge; but a recent study has begun to address this
issue~\cite{biferale2018turbulent}.

We obtain the generalized von K\'arm\'an-Howarth relation for the
stochastically forced 3D HVBK model of superfluid turbulence by using the
generating-functional approach that has been developed in
Refs.~\cite{polyakov1995turbulence,yakhot2001mean,basu2014structure}. By
carrying out direct numerical simulations (DNSs) of the 3D HVBK equations, we
show that, in the statistically steady state of homogeneous and isotropic
superfluid turbulence, the probability distribution function (PDF) $P(\gamma)$
of the ratio $\gamma$  of the magnitudes of normal-fluid and superfluid
velocities, has power-law tails that scale as $P(\gamma) \sim \gamma^3$, for
$\gamma \ll 1$, and $P(\gamma) \sim \gamma^{-3}$, for $\gamma \gg 1$; we show,
analytically, how these scaling behaviors can be understood.  Furthermore, we
show that the PDF $P(\theta)$, of the angle $\theta$ between the normal-fluid
and superfluid velocities, behaves as $P(\theta)\sim \theta$, for $\theta \ll
\theta_*$, and $P(\theta)\sim \theta^{-4}$, for $\theta_* \ll \theta \ll 1$
(with $\theta_*$ a crossover angle that we define below). We also calculate the
longitudinal-velocity structure-function exponents for both normal and
superfluid components, as a function of the temperature, to explore the
multiscaling of such structure functions in 3D HVBK superfluid turbulence.  The
parameters for our DNS runs (Table~\ref{tab:parameters}) are taken from the
measurements of Ref.~\cite{barenghi1983mfriction} on superfluid $^4$He;
therefore, our results are of direct relevance to superfluid turbulence in this
system.

The remainder of this paper is organized as follows.  Section~\ref{sec:Model}
defines the simplified version of the HVBK model and the numerical method that
we use to study superfluid turbulence in this model.  Section~\ref{sec:Results}
comprises two subsections; the first contains our analytical results for  the
analog of the von-K\'arm\'an-Howarth relation for HVBK superfluid turbulence;
the second subsection is devoted to our numerical results for the multiscaling of
HVBK structure functions and other statistical properties of HVBK turbulence.
Section~\ref{sec:Conclusions} contains a discussion of our results. Some of the
details of our calculations are given in the Appendix.
     
\section{ MODEL AND NUMERICAL SIMULATIONS}
\label{sec:Model}

We use the simplified form of the HVBK equations~\cite{roche2009HVBKdns}, which
comprise the incompressible Navier-Stokes (for the normal fluid) and Euler (for
the superfluid) equations coupled via the mutual-friction term.  In addition to
the kinematic viscosity $\nu_{\rm n}$ of the normal fluid, we include Vinen's
effective viscosity~\cite{vinen2002quantum} $\nu_{\rm s}$ in the superfluid
component to mimic the dissipation because of (a) vortex reconnections and (b)
interactions between superfluid vortices and the normal
fluid~\cite{l2007bottleneck}; $\nu_{\rm s} \ll \nu_{\rm n}$. These equations
are: 

\begin{subequations}
\begin{equation}
{\partial_{t}{\bf u}_{\rm n}} + ({\bf u}_{\rm n} \cdot \nabla) {\bf u}_{\rm n} = 
-\frac{1}{\rho_{\rm n}} \nabla p_{\rm n} +\nu_{\rm n} \nabla^2 {\bf u}_{\rm n}+
\frac{\rho_{\rm s}}{\rho}
{\bf f}_\text{mf} + {\bf f}^{\rm n}_{\rm u} ;                
\label{eq:HVBK1}
\end{equation}
\begin{equation}
\nabla \cdot {\bf u}_{\rm n} = 0;
\label{eq:HVBK2}
\end{equation}
\begin{equation} 
{\partial_{t} {\bf  u}_{\rm s}} + ({\bf u}_{\rm s} \cdot \nabla) {\bf u}_{\rm s} =
-\frac{1}{\rho_{\rm s}} \nabla p_{\rm s} +\nu_{\rm s} \nabla^2 {\bf u}_{\rm s}-\frac{\rho_{\rm n}}{\rho}
{\bf f}_\text{mf} + {\bf f}^{\rm s}_{\rm u} ; 
\label{eq:HVBK3}
\end{equation}
\begin{equation}
\nabla \cdot {\bf u}_{\rm s} = 0;
\label{eq:HVBK4}
\end{equation}
\end{subequations}
here, ${\bf u}_{\rm n} ({\bf u}_{\rm s})$, $\rho_{\rm n} (\rho_{\rm s})$,
$p_{\rm n} (p_{\rm s})$, and $ {\bf f}_{\rm u}^{\rm n} ( {\bf f}_{\rm u}^{\rm
s})$ are, respectively, the velocity, density, pressure, and external-forcing
term for the normal fluid (superfluid). The mutual-friction term
\begin{equation}
 {\bf f}_\text{mf} = \frac{B}{2}\widehat{ {\boldsymbol \omega}_{\rm s}} \times({\boldsymbol \omega}_{\rm s} 
\times ({\bf u}_{\rm n} - {\bf u}_{\rm s} )) + \frac{B'}{2}{\boldsymbol \omega}_s 
\times ( {\bf u}_{\rm n}-{\bf u}_{\rm s})
\label{eq:HVBK5}
\end{equation}
leads to energy transfer between the normal and superfluid
components~\cite{morris2008vortexlock, wacks2011STshellmodel}; $ {\bf
u}_{\rm ns} = {\bf u}_{\rm n} - {\bf u}_{\rm s} $ is the slip velocity, $ {\boldsymbol
\omega}_{\rm s} =  \nabla \times {\bf u}_{\rm s} $ is the superfluid vorticity, and $B$ and
$B'$ are the mutual-friction coefficients.

We perform extensive DNSs of the HVBK equations~(\ref{eq:HVBK1}-\ref{eq:HVBK4})
by using the pseudospectral method, with periodic boundary conditions, in a
cubical box of length $2\pi$, along each direction, and $N_{\rm c}^{3}$
collocation points; we use the $2/3$  de-aliasing rule~\cite{canuto2006} and a
constant-energy-injection scheme for forcing~\cite{lamorgese2005direct,
GanapatiNJP2011}, in which we force the Fourier modes in the first two
Fourier-space shells for the superfluid, at low temperatures, and the normal fluid, at high
temperatures. We use the second-order Adams-Bashforth scheme for time
marching~\cite{GanapatiNJP2011}. The parameters for the various runs we perform
are listed in Table~\ref{tab:parameters}.

\section{ RESULTS}\label{sec:Results}

We begin (Sec.~\ref{sec:sfhierarchy}) with our results for the
structure-function hierarchy for 3D HVBK turbulence that is statistically
steady, homogeneous, and isotropic. In particular, we obtain the hierarchy of
equations for the structure functions that are statistically steady-state
values of integer powers and products of $\Delta u_{\alpha\parallel}=[{\bf
u}_{\alpha}({\bf x}+{\bf r})-{\bf u}_{\alpha} ({\bf x})].\hat{\bf r}$ and
$\Delta u_{\alpha\perp}=[{\bf u}_{\alpha}({\bf x} +{\bf r})-{\bf
u}_{\alpha}({\bf x})]\times \hat{{\bf r}}$ ($\alpha$ can be ${\rm n}$ (normal)
or ${\rm s}$ (superfluid)), which are, respectively, velocity increments along
${\bf r}$  or perpendicular to it. We obtain explicit expressions for
third-order structure functions.  In Sec.~\ref{sec:dns}, we present results
from our DNSs of the 3D HVBK equations for the PDFs $P(\gamma)$ and
$P(\theta)$ and the longitudinal-velocity structure-function
exponents for both normal and superfluid components, as a function of
temperature; we then explore their multiscaling properties.
 
 \subsection{ Structure-Function Hierarchy}\label{sec:sfhierarchy}

We now obtain the structure-function hierarchy for normal-fluid and superfluid
velocities by using Eqs.~(\ref{eq:HVBK1}) - (\ref{eq:HVBK4}) and the external
forces ${\bf f}_{u}^n$ and ${\bf f}_{u}^s$, which are zero-mean, Gaussian
random variables with the covariances
\begin{eqnarray}
	\langle f_{{\rm u} i}^{\rm n}({\bf x},t) f_{{\rm u}j}^{\rm n}({\bf x'},t') \rangle &=& \delta(t-t')
  K^{\rm n}_{ij}({\bf x}-{\bf x'}), \nonumber \\
\langle f_{{\rm u}i}^{\rm s}({\bf x},t) f_{{\rm u}j}^{\rm s}({\bf x'},t')\rangle  &=& \delta(t-t')
  K^{\rm s}_{ij}({\bf x}-{\bf x'}),
\label{eq:covariance}
\end{eqnarray}
where both $K^{\rm n}_{ij}$ and $K^{\rm s}_{ij}$ are even functions of $({\bf x}-{\bf
x'})$, and the Cartesian indices $i, j = 1, 2, 3$. We define the two-point
generating functionals $Z$ for  ${\bf u}_{\rm n}({\bf x}_1,t_1)$ and ${\bf u}_{\rm s}({\bf
x}_1,t_1)$, ${\bf u}_{\rm n}({\bf x}_2,t_2)$ and ${\bf u}_{\rm s}({\bf x}_2,t_2)$, to
calculate the hierarchy of relations for equal-time structure function in the
nonequilibrium, statistically steady state of the stochastically forced 3D HVBK
equations.

\onecolumngrid
\noindent\rule{18cm}{0.4pt} \\
The two-point generating functional $Z$ is  
\begin{equation}
\begin{aligned}
Z(\boldsymbol{\lambda}_{1 \rm n},\boldsymbol{\lambda}_{2 \rm n},
\boldsymbol{\lambda}_{1 \rm s},\boldsymbol{\lambda}_{2 \rm s},{\bf x}_1,{\bf x}_2,t_1,t_2)
&= \langle \exp[\boldsymbol {\lambda}_{1 \rm n}\cdot{\bf u}_{\rm n}({\bf x}_1) +
\boldsymbol {\lambda}_{2 \rm n}\cdot{\bf u}_{\rm n}({\bf x}_2)+
\boldsymbol{\lambda}_{1 \rm s}
\cdot{\bf u}_{\rm s}({\bf x}_1)+ \boldsymbol {\lambda}_{2 \rm s}\cdot{\bf u}_
{\rm s}({\bf x}_2)]
\rangle = \langle Z_{\rm n} Z_{\rm s} \rangle \\
&=\int \int d{\bf x}_{1 \rm n}d{\bf x}_{2 \rm n}d{\bf x}_{1 \rm s}d{\bf x}_
{2 \rm s} P({\bf u}_{\rm n}({\bf x}_1),{\bf u}_{\rm s}({\bf x}_1),t_1;{\bf u}_
{\rm n}({\bf x}_2), {\bf u}_{\rm s}({\bf x}_2), t_2) Z_{\rm n} Z_{\rm s} ,
\label{eq:Z}
\end{aligned}	
\end{equation}
where $\boldsymbol{\lambda}_{1 \rm n},\boldsymbol{\lambda}_{2 \rm n},
\boldsymbol{\lambda}_{1 \rm s},$ and $ \boldsymbol{\lambda}_{2 \rm s}$ are the 
variables conjugate to ${\bf u}_{\rm n}({\bf x}_1), {\bf u}_{\rm n}({\bf x}_2),
{\bf u}_{\rm s}({\bf x}_1),$ and $ {\bf u}_{\rm s}({\bf x}_1),$ respectively, $
Z_{\rm n} = \exp[\boldsymbol {\lambda}_{1 \rm n}\cdot{\bf u}_{\rm n}({\bf x}_1)
+ \boldsymbol {\lambda}_{2\rm n}\cdot{\bf u}_{\rm n}({\bf x}_2)], Z_{\rm s} = 
\exp[\boldsymbol {\lambda}_{1\rm s}\cdot{\bf u}_{\rm s}({\bf x}_1) + 
\boldsymbol {\lambda}_{2 \rm s}
\cdot{\bf u}_{\rm s}({\bf x}_2)]$, and $P({\bf u}_{\rm n}({\bf x}_1), {\bf u}_{\rm s}({\bf x}_1),t_1;{\bf u}_{\rm n}({\bf x}_2), {\bf u}_{\rm s}({\bf x}_2), t_2)$
is the joint probability distribution function (JPDF) of ${\bf u}_{\rm n}$ and 
${\bf u}_{\rm s}$. 
We set $ t_1=t_2=t $, which suffices for calculating the equal-time structure functions
we consider. By taking the time derivative of Eq.~(\ref{eq:Z}), we get the 
master equations for the normal fluid and superfluid:
\begin{equation}
{\partial_t}Z\big|_{\boldsymbol{\lambda}_{1 \rm s} = \boldsymbol{\lambda}_{2 \rm s} = 0}=\big\langle\big[\boldsymbol{\lambda}_{1 \rm n}\cdot{{\partial_t}
\bf u}_{\rm n}({\bf x}_1) + \boldsymbol{\lambda}_{2 \rm n}\cdot{{\partial_t}
\bf u}_{\rm n}({\bf x}_2) + \boldsymbol{\lambda}_{1 \rm s}\cdot{{\partial_t}
\bf u}_{\rm s}({\bf x}_1) + \boldsymbol{\lambda}_{2 \rm s}\cdot{{\partial_t}
\bf u}_{\rm s}({\bf x}_2)\big]Z_{\rm n}Z_{\rm s}\big \rangle\big|_{\boldsymbol{\lambda}_{1 \rm s} = \boldsymbol{\lambda}_{2 \rm s} = 0};
\label{eq:DtZn}
\end{equation}
\begin{equation}
{\partial_t}Z\big|_{\boldsymbol{\lambda}_{1 \rm n} = \boldsymbol{\lambda}_{2 \rm n} = 0}=\big\langle\big[\boldsymbol{\lambda}_{1 \rm n}\cdot{{\partial_t}
\bf u}_{\rm n}({\bf x}_1) + \boldsymbol{\lambda}_{2 \rm n}\cdot{{\partial_t}
\bf u}_{\rm n}({\bf x}_2) + \boldsymbol{\lambda}_{1 \rm s}\cdot{{\partial_t}
\bf u}_{\rm s}({\bf x}_1) + \boldsymbol{\lambda}_{2 \rm s}\cdot{{\partial_t}
\bf u}_{\rm s}({\bf x}_2)\big]Z_{\rm n}Z_{\rm s}\big\rangle\big|_{\boldsymbol{\lambda}_{1
\rm n} = \boldsymbol{\lambda}_{2 \rm n} = 0};
\label{eq:DtZs}
\end{equation}
by substituting Eqs.~(\ref{eq:HVBK1} - \ref{eq:covariance}) in  Eq.~(\ref{eq:DtZn}) and Eq.~(\ref{eq:DtZs}) 
we get, in the statistically steady state,
\begin{align}
& \bigg \langle \bigg[\frac{\partial^2 Z_{\rm n}}{\partial r_{i}\partial
\lambda_{1 {\rm n}i}}\bigg] \bigg \rangle +\bigg \langle \bigg[\frac{\partial^2
Z_{\rm n}}{\partial r_{i}\partial \lambda_{2{\rm n}i}}\bigg] \bigg \rangle 
+\frac{\rho_{\rm s}}{\rho} \bigg \langle \Big[{\boldsymbol{\lambda}_{1 \rm n}
\cdot {\bf f}_\text{mf} ({\bf x}_1) +  \boldsymbol{\lambda}_{2 \rm n} \cdot 
{\bf f}_\text{mf}({\bf x}_2)} \Big] Z_{ \rm n}  \bigg \rangle=I^{\rm n}_{\rm p}
+I^{\rm n}_{\rm f} + D_{\rm n} ,
\label{eq:Steady_n}
\end{align}
\begin{align}
& \bigg \langle \bigg [\frac{\partial^2 Z_{\rm s}} {\partial
r_{i}\partial\lambda_{1{\rm s}i}}\bigg ] \bigg \rangle + \bigg \langle 
\bigg [ \frac{\partial^2 Z_{\rm s}}
{\partial r_{i}\partial\lambda_{2{\rm s}i}}\bigg ] \bigg \rangle 
-\frac{\rho_{\rm n}}{\rho}\bigg \langle \Big[{\boldsymbol{\lambda}_{1 \rm s} 
\cdot {\bf f}_\text{mf} ({\bf x}_1) + \boldsymbol{\lambda}_{2 \rm s} \cdot
{\bf f}_\text{mf}({\bf x}_2)}\Big] Z_{\rm s} \bigg \rangle = I^{\rm s}_{\rm p}
+I^{\rm s}_{\rm f}+D_{\rm s} ,
\label{eq:Steady_s}
\end{align}
where $r_i \; (i = 1, 2, 3)$ are the Cartesian components of the relative vector 
${\bf r}= ({\bf x}_1-{\bf x}_2)$, with $r = |{\bf r}|$ and $\hat{{\bf r}} = 
{\bf r}/r$, and $I^{\rm n}_{\rm p} (I^{\rm s}_{\rm p}), I^{\rm n}_{\rm f}
(I^{\rm s}_{\rm f})$, and $D_{\rm n} (D_{\rm s})$, 
which arise, respectively, from the pressure, forcing, and dissipation terms 
from the normal fluid (superfluid), are defined as follows: 
\begin{eqnarray}
I^{\rm n}_p & =& -\Big \langle \Big [\boldsymbol{\lambda}_{1 \rm n}\cdot
\frac{1}{\rho_{\rm n}} \nabla p_{\rm n}({\bf x}_1)+\boldsymbol{\lambda}_{2 \rm n} \cdot \frac{1}{\rho_{\rm n}} \nabla p_{\rm n}({\bf x}_2)  \Big ] Z_{\rm n} 
\Big \rangle ; \nonumber \\
I^{\rm s}_p & =&-\Big \langle\Big[\boldsymbol{\lambda}_{1 \rm s} \cdot \frac{1} {\rho_{\rm s}} \nabla p_{\rm s}({\bf x}_1) + \boldsymbol{\lambda}_{2 \rm s}
\cdot \frac{1}{\rho_{\rm s}} \nabla p_{\rm s}({\bf x}_2) \Big]Z_{s}\Big 
\rangle ; \nonumber \\
I^{\rm n}_{\rm f} &=& \Big \langle \Big [\boldsymbol{\lambda}_{1 \rm n} \cdot
{\bf f}_{\rm u}^{\rm n} ({\bf x}_1)+ \boldsymbol{\lambda}_{2 \rm n} \cdot
{\bf f}_{\rm u}^{\rm n}({\bf x}_2) + \Big]Z_{\rm n}  \Big \rangle; \nonumber\\ 
I^{\rm s}_{\rm f} &=& \Big \langle \Big [\boldsymbol{\lambda}_{1\rm s} \cdot 
{\bf f}_{\rm u}^{\rm n} ({\bf x}_1)+ \boldsymbol{\lambda}_{2 \rm s} \cdot 
{\bf f}_{\rm u}^{\rm n}({\bf x}_2) + \Big] Z_{\rm s} \Big \rangle; \nonumber \\ 
D_{\rm n} &=&\Big \langle \Big [\nu_{\rm n} \Big(\boldsymbol{\lambda}_{1 \rm n}
\cdot \nabla^2 {\bf u}_{\rm n}({\bf x}_1) + \boldsymbol{\lambda}_{2 \rm n} 
\cdot \nabla^2 {\bf u}_{\rm n} ({\bf x}_2)\Big) \Big] Z_{\rm n} \Big \rangle;
\nonumber \\ 
D_{\rm s} &=&\Big \langle \Big [\nu_{\rm s} \Big (\boldsymbol{\lambda}_{1\rm s}
\cdot \nabla^2 {\bf u}_{\rm s}({\bf x}_1) + \boldsymbol{\lambda}_{2 \rm s} 
\cdot \nabla^2 {\bf u}_{\rm s}({\bf x}_2)\Big)\Big] Z_{\rm s} \Big \rangle . 
\label{eq:IpIfD1}
\end{eqnarray}
It is useful to define ${\bf x}$, the center-of-mass coordinate; clearly ${\bf
x}_1 = {\bf x} + \frac{\bf r}{2}$ and $ {\bf x}_2 = {\bf x} - \frac{\bf r}{2}$.
Equations~(\ref{eq:Steady_n}) and ~(\ref{eq:Steady_s}) are invariant under the
Galilean transformation ${\bf r}'={\bf r}-{\bf u}_0 t$, $t' = t$, and ${\bf
u}'_{\alpha} = {\bf u}_ {\alpha} + {\bf u}_0$; here, $\alpha$ stands for n and
s, with ${\bf u}_0$ a constant velocity. If we impose the homogeneity condition
$\frac{\partial Z}{\partial \bf x} = 0$, we find that $Z$ depends only on $\bf
r$.

For simplicity, we consider $\boldsymbol{\lambda}_{1 \rm n}$ antiparallel to
$\boldsymbol{\lambda}_{2 \rm n}$, i.e., $\boldsymbol{\lambda}_{1 \rm n} =
-\boldsymbol {\lambda}_{2n \rm } \equiv \boldsymbol{\lambda}_{\rm n}$ and
$\boldsymbol{\lambda}_{1 \rm s}$ antiparallel to $\boldsymbol{\lambda}_{2 \rm
s}$, i.e., $\boldsymbol{\lambda}_{1 \rm s} = -\boldsymbol{\lambda}_{2\rm s}
\equiv \boldsymbol{\lambda}_{\rm s}$. (For a discussion of this choice, see 
footnote [47] of Ref.~\cite{basu2014structure} for the formally
related problem of MHD turbulence.) We get the generalized structure function
$\langle (\Delta u^m_{{\rm n}i}) (\Delta u^n_{{\rm s}i}) \rangle$ by taking the
order $m$ derivative of $Z$ with respect to the Cartesian component
$\boldsymbol{\lambda}_{{\rm n} i}$ and the order $n$ derivative of $Z$ with
respect to the Cartesian component $\boldsymbol{\lambda}_{{\rm s} i}$.  In the
case of homogeneous and isotropic 3D HVBK superfluid turbulence, $Z_n$ depends on
$\eta_{n1} = r$, $\eta_{{\rm n}2} = \boldsymbol {\lambda}_{\rm n} \cdot
\hat{{\bf r}}_{\rm n} = \boldsymbol{\lambda}_{\rm n} \cos{\theta_{\rm n}}$, and
$\eta_{{\rm n}3} = \boldsymbol{\lambda}_{\rm n} \sin{\theta_{\rm n}}$ and
$Z_{\rm s}$  depends on $\eta_{{\rm s}1} = r$, $\eta_{{\rm s}2}= \boldsymbol
{\lambda}_{\rm s} \cdot \hat{{\bf r}}_{ \rm s} = \boldsymbol{\lambda}_{\rm s}
\cos{\theta_{ \rm s}}$ and $\eta_{{\rm s} 3} = \boldsymbol{\lambda}_{\rm s}
\sin{\theta_{\rm s}}$.  In terms of these variables the generating functionals
can be written as follows:
\begin{equation}
Z_{\rm n}=\exp[\eta_{{\rm n}2}\Delta u_{{\rm n}\parallel}+\eta_{{\rm n}3}
\Delta u_{{\rm n}\perp}];\hspace{2cm}
Z_{\rm s} = \exp[\eta_{{\rm s}2}\Delta u_{{\rm s} \parallel}+ \eta_{{\rm s}3}
\Delta u_{{\rm s}\perp}];  
\label{eq:ZnZs}
\end{equation}
here, $\Delta u_{\alpha\parallel}=[{\bf u}_{\alpha}({\bf x}+{\bf r})-{\bf\alpha}
({\bf x})].\hat{\bf r}$ and $\Delta u_{\alpha\perp}=[{\bf u}_{\alpha}({\bf x}
+{\bf r})-{\bf u}_{\alpha}({\bf x})]\times \hat{{\bf r}}$ ($\alpha$ can be n
(normal) or s (superfluid)) are, respectively, velocity increments along
${\bf r}$  or perpendicular to it; similar increments can be defined for 
the forcing and mutual-friction terms. By using the variables $r,
\eta_{{\rm n}2}, \eta_{{\rm n}3}, \eta_{{\rm s}2}$ and $\eta_{{\rm s}3}$ in 
Eqs.~(\ref{eq:Steady_n}) and ~(\ref{eq:Steady_s}), in the statistically steady state, 
we get:
\begin{align}
&\Big \langle \Big[\partial_{r} \partial_{\eta_{{\rm n}2}}+\frac{2}{r} 
\partial_{\eta_{{\rm n}2}}-\frac{1}{r} \frac{\eta_{{\rm n}2}}{\eta_{{\rm n}3}}
\partial_{\eta_{{\rm n}3}} + \frac{\eta_{{\rm n}3}}{r}\partial_{\eta_{{\rm n}2}}\partial_{\eta_{{\rm n}3}} -\frac{\eta_{{\rm n}2}} {r}{\partial^2_{\eta^2_
{{\rm n}3}}}\Big]Z_{\rm n}  \Big \rangle + \frac{\rho_{\rm s}}{\rho}\Big
\langle \Big[ \eta_{{\rm n}2} \Delta f_{\text{mf}_{\parallel}}+\eta_{{\rm n}3}
\Delta f_{\text{mf}_{\perp}} \Big] Z_{\rm n} \Big \rangle =I^{\rm n}_{\rm p}
+I^{\rm n}_{\rm f}+D_{\rm n} ;
\label{eq:men}
\end{align}
\begin{align}
&\Big \langle \Big[\partial_{r} \partial_{\eta_{{\rm s}2}}+\frac{2}{r} 
\partial_{\eta_{{\rm s}2}}-\frac{1}{r} \frac{\eta_{{\rm s}2}}{\eta_{{\rm s}3}}
\partial_{\eta_{{\rm s}3}} +  \frac{\eta_{{\rm s}3}}{r}\partial_{\eta_{{\rm s}2}}\partial_{\eta_{{\rm s}3}} -\frac{\eta_{{\rm s}2}} {r}{\partial^2_{\eta^2_
{{\rm s}3}}}\Big]Z_{\rm s}  \Big \rangle - \frac{\rho_{\rm n}}{\rho}\Big
\langle \Big[ \eta_{{\rm s}2} \Delta f_{\text{mf}_{\parallel}}+\eta_{{\rm s}3}
\Delta f_{\text{mf}_{\perp}} \Big] Z_{\rm s} \Big \rangle = I^{\rm s}_{\rm p}
+ I^{\rm s}_{\rm f} +D_{\rm s} . 
\label{eq:mes}
\end{align}
If we multiply Eq.(\ref{eq:men}) by $\eta_{{\rm n}3}$ and Eq.(\ref{eq:mes}) by
$\eta_{{\rm s}3}$, and we substitute Eq.(\ref{eq:ZnZs}) in
Eqs.(\ref{eq:men}-\ref{eq:mes}), we obtain, after some simplification: 
\begin{align}
& \bigg \langle \eta_{{\rm n}3} \bigg[\frac{\partial \Delta u_{{\rm n}_
\parallel}} {\partial r} + \Delta u_{{\rm n}_\parallel}\bigg(\eta_{{\rm n}2}
\frac{\partial \Delta u_{{\rm n}_\parallel}}{\partial r}+ \eta_{{\rm n}3}
\frac{\partial \Delta u_{{\rm n}_\perp}}{\partial r}\bigg)+ 
\frac{2}{r}\Delta u_{{\rm n}_\parallel}- \frac{1}{r}\frac{\eta_{{\rm n}2}}
{\eta_{{\rm n}3}}\Delta u_{{\rm n}_\perp} +\frac{\eta_{{\rm n}3}}{r}\Delta
u_{{\rm n}_\parallel} \Delta u_{{\rm n}_\perp} - \frac{\eta_{{\rm n}2}}{r}
(\Delta u_{{\rm n}_\perp})^2 \bigg] Z_{\rm n} \bigg \rangle  \nonumber  \\ & + 
\frac{\rho_{\rm s}}{\rho}\Big \langle \eta_{{\rm n}3}\Big[ \eta_{{\rm n}2} 
\Delta f_{\text{mf}_{\parallel}}+\eta_{{\rm n}3}\Delta f_{\text{mf}_{\perp}} 
\Big] Z_{\rm n} \Big \rangle = \eta_{{\rm n}3}\Big (I^{\rm n}_{\rm p} + 
I^{\rm n}_{\rm f}+D_{\rm n} \Big ) ;
\label{eq:men1}
\end{align}

\begin{align}
& \bigg \langle \eta_{{\rm s}3} \bigg[\frac{\partial \Delta u_{{\rm s}_
\parallel}} {\partial r} + \Delta u_{{\rm s}_\parallel}\bigg(\eta_{{\rm s}2}
\frac{\partial \Delta u_{{\rm s}_\parallel}}{\partial r}+ \eta_{{\rm s}3}
\frac{\partial \Delta u_{{\rm s}_\perp}}{\partial r}\bigg)+ \frac{2}{r}\Delta
u_{{\rm s}_\parallel}- \frac{1}{r}\frac{\eta_{{\rm s}2}}{\eta_{{\rm s}3}}
\Delta u_{{\rm s}_\perp} +\frac{\eta_{{\rm s}3}}{r}\Delta u_{{\rm s}_\parallel} \Delta u_{{\rm s}_\perp} - \frac{\eta_{{ \rm s}2}}{r} (\Delta u_{{\rm s}_
\perp})^2 \bigg] Z_{\rm s} \bigg \rangle  \nonumber  \\ & - \frac{\rho_{\rm n}}
{\rho}\Big \langle \eta_{{\rm s}3}\Big[ \eta_{{\rm s}2} \Delta f_{\text{mf}_
{\parallel}}+\eta_{{\rm s}3}\Delta f_{\text{mf}_{\perp}} \Big] Z_{\rm s} \Big
\rangle = \eta_{{\rm s}3}\Big (I^{\rm s}_{\rm p}+I^{\rm s}_{\rm f} + D_{\rm s} \Big ) .
\label{eq:mes1}
\end{align}

The pressure contributions, $I_{\rm p}^{\rm n}$ and $I_{\rm p}^{\rm s}$,
vanish, as in the case of homogeneous, isotropic fluid
turbulence~\cite{yakhot2001mean}, if we consider only third-order structure
functions. This follows from the symmetries of the velocity and pressure fields
under spatial inversion (Appendix).

The forcing contributions, $I_{\rm f}^{\rm n}$ and $I_{\rm f}^{\rm s}$, can also be 
neglected in the inertial range of scales in 3D HVBK superfluid turbulence (see below); 
these can be written as follows:
\begin{eqnarray}
I^{\rm n}_{\rm f}&=&\Big \langle \Big [\boldsymbol{\lambda}_{\rm n} \cdot
{\bf f}_{\rm u}^{\rm n} ({\bf x}_1)- \boldsymbol{\lambda}_{\rm n} \cdot 
{\bf f}_{\rm u}^{\rm n}({\bf x}_2)\Big] Z_{\rm n} \Big \rangle \equiv \Big 
\langle \Big [{\eta}_{{\rm n}2}{\Delta f}_{{\rm u}_\parallel}^{\rm n} + 
{\eta}_{{\rm n}3}{\Delta f}_{{\rm u}_\perp}^{\rm n}  \Big]Z_{\rm n}\Big\rangle ;
\end{eqnarray}
\begin{eqnarray}
I^{\rm s}_{\rm f}&=&\Big \langle \Big [\boldsymbol{\lambda}_{\rm s} \cdot 
{\bf f}_{\rm u}^{\rm s} ({\bf x}_1)- \boldsymbol{\lambda}_{\rm s} \cdot 
{\bf f}_{\rm u}^{\rm s}({\bf x}_2)\Big] Z_{\rm s} \Big \rangle \equiv \Big 
\langle \Big [{\eta}_{{\rm s}2}{\Delta f}_{{\rm u}_\parallel}^{\rm s} + 
 {\eta}_{{\rm s}3}{\Delta f}_{{\rm u}_\perp}^{\rm s} \Big]Z_{\rm s}\Big\rangle .
\end{eqnarray}
If we now use the Furutsu-Novikov-Donsker formula~\cite{woyczynski1998,McComb1990} 
we get, after some simplification: 
\begin{align}
I^{\rm n}_{\rm f}= & \Big \langle \Big[ \eta_{{\rm n}2}^2\Big(K_{{\rm n}_
{\parallel\parallel}}(0)-K_{{\rm n}_{\parallel \parallel}} ({\bf r})\Big)+
2\eta_{{\rm n}2}\eta_{{\rm n}3}\Big(K_{{\rm n}_{\parallel\perp}}(0)-
K_{{\rm n}_{\parallel\perp}}({\bf r})\Big)+\eta_{{\rm n}3}^2\Big(K_{{\rm n}_
{\perp\perp}}(0)- K_{{\rm n}_{\perp\perp}}({\bf r})\Big) \Big] Z_{\rm n} \Big\rangle ;
\label{eq:IfFinal_n}
\end{align}
\begin{align}
I^{\rm s}_{\rm f} = & \Big \langle \Big [\eta_{{\rm s}2}^2\Big(K_{{\rm s}_
{\parallel\parallel}}(0)-K_{{\rm s}_{\parallel\parallel}}({\bf r})\Big)+ 
2\eta_{{\rm s}2}\eta_{{\rm s}3}\Big(K_{{\rm s}_{\parallel\perp}}(0)-
K_{{\rm s}_{\parallel\perp}}({\bf r})\Big)+\eta_{{\rm s}3}^2\Big(K_{{\rm s}_
{\perp\perp}}(0)- K_{{\rm s}_{\perp\perp}}({\bf r})\Big) \Big]  Z_{\rm s} 
\Big\rangle.
\label{eq:IfFinal_s}
\end{align}
These terms contribute to the relations between third-order structure functions
only at $\mathcal{O}((r/r_{\rm f})^2)$,  where $r_{\rm f}$ is the forcing
length scale, so we can neglect them in the inertial range, for $r\ll r_{\rm
f}$, in the case of 3D HVBK superfluid turbulence (see the discussion below Eq. (7) in
Ref.~\cite{yakhot2001mean} for the case of classical-fluid turbulence in 3D).

The dissipation terms are:
\begin{eqnarray}
D_{\rm n}=\nu_{\rm n} \Big \langle \Big[{\boldsymbol\lambda}_{\rm n}\cdot 
\nabla^2_{{\bf x}_1}{\bf u}_{\rm n}{({\bf x}_1)}-{\boldsymbol \lambda}_{\rm n}
\cdot\nabla^2_{{\bf x}_2} {\bf u}_{\rm n}{({\bf x}_2)}\Big] Z_{\rm n} \Big
\rangle; \hspace{1cm} 
D_{\rm s} = \nu_{\rm s} \Big \langle  \Big [{\boldsymbol \lambda}_{\rm s}\cdot
\nabla^2_{{\bf x}_1}{\bf u}_{\rm s}{({\bf x}_1)}- {\boldsymbol \lambda}_{\rm s}
\cdot\nabla^2_{{\bf x}_2}{\bf u}_{\rm s}{({\bf x}_2)}\Big] Z_{\rm s} \Big 
\rangle .
\label{eq:DSemiFinal}
\end{eqnarray}
If we take the limit of large Reynolds number, i.e., $\nu_{\rm n} \rightarrow 
0$ and $\nu_{\rm s} \rightarrow 0$, define $\epsilon_{{\rm n}_\parallel}
({\bf x}_1)+\epsilon_{{\rm n}_\parallel}({\bf x}_2)= \epsilon_{{\rm n}_\parallel}$, $ \epsilon_{{\rm n}_\perp}({\bf x}_1)+\epsilon_{{\rm n}_\perp}
({\bf x}_2)=\epsilon_{{\rm n}_\perp}$, $\epsilon_{{\rm s}_\parallel}({\bf x}_1)
+ \epsilon_{{\rm s}_\parallel}({\bf x}_2)= \epsilon_{{\rm s}_\parallel}$, and
$\epsilon_{{\rm s}_\perp}({\bf x}_1)+\epsilon_{{\rm s}_\perp}({\bf x}_2) = 
\epsilon_{{\rm s}_\perp}$ we can simplify Eq.~(\ref{eq:DSemiFinal}) (see the
Appendix for details) to get:  
\begin{align}
-D_{\rm n}= & \Big \langle \Big [\eta_{{\rm n}2}^2\epsilon_{{\rm n}_\parallel}+
\eta_{{\rm n}3}^2 \epsilon_{{\rm n}_\perp} \Big ]Z_{\rm n}  \Big \rangle
+ 2 \Big \langle \eta_{{\rm n}2}\eta_{{\rm n}3} \Big [\Big(\epsilon_{{\rm n}_
\parallel}{({\bf x}_1)} \epsilon_{{\rm n}_\perp}{({\bf x}_1)}\Big)^{\frac{1}{2}}
 + \Big(\epsilon_{{\rm n}_\parallel}{({\bf x}_2)}\epsilon_{{\rm n}_\perp}
{({\bf x}_2)}\Big)^{\frac{1}{2}}\Big] Z_{\rm n} \Big \rangle;
\label{eq:DFinal}
\end{align}
\begin{align}
-D_{\rm s}= & \Big \langle \Big [\eta_{{\rm s}2}^2\epsilon_{{\rm s}_\parallel}+
\eta_{{\rm s}3}^2 \epsilon_{{\rm s}_\perp} \Big ]Z_{\rm s}  \Big \rangle
+ 2 \Big \langle \eta_{{\rm s}2}\eta_{{\rm s}3} \Big [\Big(\epsilon_{{\rm s}_
\parallel}{({\bf x}_1)} \epsilon_{{\rm s}_\perp}{({\bf x}_1)}\Big)^{\frac{1}{2}}
+ \Big(\epsilon_{{\rm s}_\parallel}{({\bf x}_2)}\epsilon_{{\rm s}_\perp}
{({\bf x}_2)}\Big)^{\frac{1}{2}}\Big] Z_{\rm s} \Big \rangle .
\label{eq:DFinal}
\end{align}
If we take the derivative $\partial_{\eta_{{\rm n}2}}^2 \partial_{\eta_{{\rm n}3}}$ of  Eq.~(\ref{eq:men1}) and the limits
$\eta_{{\rm n}2},\eta_{{\rm n}3} \rightarrow 0$, we get
\begin{equation}
\frac{\partial \langle(\Delta u_{{\rm n}_\parallel})^3 \rangle}{\partial r}+
\frac{2}{r}\langle(\Delta u_{{\rm n}_\parallel})^3 \rangle -\frac{4}{r} 
\langle \Delta u_{{\rm n}_\parallel} (\Delta u_{{\rm n}_\perp})^2
\rangle+2\frac{\rho_{\rm s}}{\rho}\langle \Delta f_\text{mf$_\parallel$}
\Delta u_{{\rm n}_\parallel}\rangle = -2 \langle \epsilon_{{\rm n}_\parallel}
\rangle ;
\label{eq:KH1}
\end{equation}
the derivative $\partial_{\eta_{{\rm n}3}}^3$ of Eq.~(\ref{eq:men1}) yields, in the 
limits $\eta_{{\rm n}2},\eta_{{\rm n}3} \rightarrow 0$,
\begin{equation}
\frac{\partial \langle \Delta u_{{\rm n}_\parallel}(\Delta u_{{\rm n}_\perp})^2
\rangle}{\partial r} +\frac{4}{r} \langle \Delta u_{{\rm n}_\parallel} 
(\Delta u_{{\rm n}_\perp})^2 \rangle + 2\frac{\rho_{\rm s}}{\rho}\langle
\Delta f_\text{mf$_\perp$} \Delta u_{{\rm n}_\perp}\rangle = -2 \langle 
\epsilon_{{\rm n}_\perp} \rangle .
\label{eq:KH2}
\end{equation}
From the derivative $\partial_{\eta_{{\rm s}2}}^2 \partial_{\eta_{{\rm s}3}}$ of
Eq.~(\ref{eq:mes1}), we obtain, in the
limits $\eta_{{\rm s}2},\eta_{{\rm s}3} \rightarrow 0$,  
\begin{equation}
\frac{\partial \langle(\Delta u_{{\rm s}_\parallel})^3 \rangle}{\partial r}+
\frac{2}{r}\langle(\Delta u_{{\rm s}_\parallel})^3 \rangle -\frac{4}{r} 
\langle \Delta u_{{\rm s}_\parallel} (\Delta u_{{\rm s}_\perp})^2 \rangle - 
2\frac{\rho_{\rm n}}{\rho}\langle \Delta f_\text{mf$_\parallel$}
\Delta u_{{\rm s}_\parallel}\rangle = -2 \langle \epsilon_{{\rm s}_\parallel}
\rangle ;
\label{eq:KH3}
\end{equation}
similarly, the derivative $\partial_{\eta_{{\rm s}3}}^3$ of Eq.~(\ref{eq:mes1}) gives,
in the limits $\eta_{{\rm s}2},\eta_{{\rm s}3} \rightarrow 0$, 
\begin{equation}
\frac{\partial \langle \Delta u_{{\rm s}_\parallel}(\Delta u_{{\rm s}_\perp})^2
\rangle}{\partial r} +\frac{4}{r} \langle \Delta u_{{\rm s}_\parallel} 
(\Delta u_{{\rm s}_\perp})^2 \rangle - 2\frac{\rho_{\rm n}}{\rho}\langle 
\Delta f_\text{mf$_\perp$} \Delta u_{{\rm s}_\perp}\rangle = -2 \langle 
\epsilon_{{\rm s}_\perp} \rangle .
\label{eq:KH4}
\end{equation}

Equations~(\ref{eq:KH1})-(\ref{eq:KH4}) are the (3D HVBK, statistically
homogeneous, isotropic superfluid turbulence) analogs of the von K\'arm\'an-Howarth
relation for statistically homogeneous and isotropic fluid turbulence.  If we
make the simplifying assumption (as in Ref.~\cite{biferale2018turbulent}) that
the mutual friction is not significant in the inertial range of scales, then we
find the usual von K\'arm\'an-Howarth relation, as in conventional
classical-fluid turbulence. However, numerical simulations (see the next
subsection~\ref{sec:dns} for our results, Eqs. (11d)-(11f) and Figs. 3(d)-3(f)
in Ref.~\cite{biferale2018turbulent}, and, for 2D HVBK turbulence, Fig. 3 (f) of
Ref.~\cite{shukla2015homogeneous}) indicate that the mutual-friction
contribution is non-negligible in the inertial range of scales. Therefore, we
must retain it in the structure-function hierarchy as we have done in
Eqs.~(\ref{eq:KH1})-(\ref{eq:KH4}). Note that, if there is complete alignment
of the normal and superfluid velocities in the statistically steady state, then
the mutual-friction term can be neglected; however, as we show in
subsection~\ref{sec:dns}, this alignment is imperfect.

We note, in passing, that we can also develop a structure-function hierarchy
for the case of statistically steady, homogeneous, isotropic 2D HVBK
superfluid turbulence~\cite{shukla2015homogeneous,pandit2017overview} by using the
generating-functional methods we have outlined above for 3D HVBK superfluid turbulence. In
this 2D case, we must distinguish between \textit{forward- and inverse-cascade
regimes}~\cite{shukla2015homogeneous,pandit2017overview}; in the former, there
is a forward cascade of enstrophy, from the forcing length scale to smaller
length scales; in the latter, there is an inverse cascade of energy towards
large length scales.  If we recall that there is no dissipative anomaly in the
forward-cascade regime in 2D turbulence~\cite{pandit2017overview}, we see
immediately that we obtain Eqs.~(\ref{eq:KH1})-(\ref{eq:KH4}) with the
dissipation terms on the right-hand side set to zero.  In the inverse-cascade
regime, the forcing contribution does not vanish, but it is of
$\mathcal{O}(1)$, because $r \gg r_f$. Therefore, in the inverse-cascade
regime, the right-hand sides (RHSs) of Eqs.~(\ref{eq:KH1})-(\ref{eq:KH4}) do
not have dissipation terms (like $-2 \langle \epsilon_{{\rm n}_\parallel}
\rangle$); instead, the RHSs of Eqs.~(\ref{eq:KH1})-(\ref{eq:KH4}) are $[K^{\rm
n}_{ij}(0) - K^{\rm n}_{ij}(r)]$ and $[K^{\rm s}_{ij}(0) - K^{\rm s}_{ij}(r)]$,
where the argument $0$ indicates zero spatial separation in the force
covariances~(\ref{eq:covariance}). For $r/r_{\rm f} \gg 1$ (of relevance to
the inverse-cascade regime), $  K^{\rm n}_{ij}(r),  K^{\rm s}_{ij}(r) \to 0$,
so we only have $ K^{\rm n}_{ij}(0)$ or  $K^{\rm s}_{ij}(0)$ on the RHSs  of
Eqs.~(\ref{eq:KH1})-(\ref{eq:KH4}); these are positive constants, clear
signatures of an inverse cascade.

\noindent\rule{18cm}{0.4pt} \\

\twocolumngrid

\begin{table*}
\begin{tabular}{c c  c  c  c  c  c  c  c  c  c  c  c  c  c  c  c  c  c}
\hline
Run & $N_{\rm c}$ & $T$ & $\rho_{\rm n}/\rho $ & $B$ & $ B'$ & $\nu_{\rm n} $ &
$\nu_{\rm s}$ & $dt$ & $f_{\rm n} $& $f_{\rm s}$ & $ \lambda_{\rm n} $ & 
$ \lambda_{\rm s} $ &  $ Re_\lambda^{\rm n} $ & $  Re_\lambda^{\rm s} $ & $
T_{\rm{eddy}}^{\rm n}  $ & $  T_{\rm{eddy}}^{\rm s} $ & $ k_{\rm{max}} \eta_{\rm n} $ & $ 
k_{\rm {max}} \eta_{\rm s}$ \\ 
\hline\hline
\bf R1 & $ 512$ & $1.30$  & $ 0.045 $   & $ 1.526  $   & $0.616  $  &
$ 1\times10^{-3}  $    & $ 1\times10^{-4}  $  & $
9\times 10^{-4} $  & $ 0.00 $ & $ 0.03 $  & $ 0.076 $ & $
0.044 $  & $ 30 $ & $ 183 $  & $ 1.22 $ & $ 1.15 $ & $ 1.79 $
& $ 0.43 $ \\
\hline

\bf R2 & $ 512 $ & $1.50$ & $ 0.111 $   & $ 1.296  $   & $ 0.317  $
& $ 1\times10^{-3}    $ & $ 1\times10^{-4}  $  & $
9\times 10^{-4} $  & $ 0.00 $ & $ 0.03 $  & $ 0.086  $ & $
0.050  $  & $ 33 $ & $ 204 $  & $ 1.31 $ & $ 1.22 $ & $ 1.92
$ & $ 0.46$ \\
\hline

\bf R3 & $ 512 $ & $1.70$  & $ 0.229 $   & $ 1.10  $   & $0.107  $  &
$ 1\times10^{-3}    $    & $ 1\times10^{-4}  $  & $
9\times 10^{-4} $  & $ 0.00 $ & $ 0.03 $  & $ 0.096  $ & $
0.059  $  & $ 34 $ & $ 225 $  & $ 1.36 $ & $ 1.22 $ & $ 2.09
$ & $ 0.51 $ \\
\hline

\bf R4 & $ 512 $ & $1.80$ & $ 0.313 $   & $ 1.024  $   & $0.052  $  &
$ 1\times10^{-3}    $    & $ 1\times10^{-4}  $  & $
9\times 10^{-4} $  & $ 0.00 $ & $ 0.03 $  & $ 0.102  $ & $
0.064  $  & $ 34 $ & $ 236 $  & $ 1.30 $ & $ 1.21 $ & $ 2.23
$ & $ 0.55 $ \\
\hline

\bf R5 & $ 512 $& $1.85$  & $ 0.364 $   & $ 0.996  $   & $0.041  $  &
$ 1\times10^{-3}    $    & $ 1\times10^{-4}  $  & $
9\times 10^{-4} $  & $ 0.00 $ & $ 0.03 $  & $ 0.107  $ & $
0.069  $  & $ 36 $ & $ 250 $  & $ 1.38 $ & $ 1.28 $ & $ 2.33
$ & $ 0.57 $ \\
\hline

\bf R6 & $ 512 $& $1.90$  & $ 0.420 $   & $ 0.98  $   & $0.04  $  & $
1\times10^{-3}    $    & $ 1\times10^{-4}  $  & $
9\times 10^{-4} $  & $ 0.00 $ & $ 0.03 $  & $ 0.114  $ & $
0.075 $  & $ 37 $ & $ 265 $  & $ 1.53 $ & $ 1.40$ & $ 2.43 $
& $ 0.60 $ \\
\hline

\bf R7 & $ 512 $& $2.10$  & $ 0.741 $   & $ 1.298  $   & $ -0.065  $
& $ 1\times10^{-3}    $    & $ 1\times10^{-4}  $  & $
9\times 10^{-4} $  & $ 0.04 $ & $ 0.00 $  & $ 0.121  $ & $
0.083 $  & $ 49 $ & $ 332 $  & $ 1.24 $ & $ 1.17 $ & $ 2.23 $
& $ 0.59 $ \\
\hline

\bf R8 & $ 512 $& $2.17$  & $ 0.95 $   & $ 3.154  $   & $ -1.272  $
& $ 1 \times10^{-3}    $    & $ 1\times10^{-4}  $  & $
9\times 10^{-4} $  & $ 0.04 $ & $ 0.00 $  & $ 0.121  $ & $
0.095 $  & $ 53 $ & $ 421 $  & $ 1.12 $ & $ 1.08$ & $ 2.15 $
& $ 0.60 $ \\
\hline

\bf R9 & $ 512 $& $1.30$  & $ 0.045 $   & $ 1.526  $   & $ 0.616  $
& $ 2.32 \times10^{-3}    $    & $ 0.1\times10^{-3}  $  & $
9\times 10^{-4} $  & $ 0.03 $ & $ 0.03 $  & $ 0.062  $ & $
0.038 $  & $ 39 $ & $ 185 $  & $ 1.19 $ & $ 1.11$ & $ 1.82 $
& $ 0.50 $ \\
\hline

\bf R10 & $ 512 $ & $1.50$ & $ 0.111 $   & $ 1.296  $   & $ 0.317  $
& $ 0.83\times10^{-3}    $ & $ 0.17\times10^{-3}  $  & $
9\times 10^{-4} $  & $ 0.03 $ & $ 0.03 $  & $ 0.084  $ & $
0.058  $  & $ 41 $ & $ 144 $  & $ 0.79 $ & $ 0.75 $ & $ 1.68
$ & $ 0.63$ \\
\hline

\bf R11 & $ 512 $ & $1.70$  & $ 0.229 $   & $ 1.10  $   & $0.107  $  &
$ 0.39\times10^{-3}    $    & $ 0.232\times10^{-3}  $  & $
9\times 10^{-4} $  & $ 0.03 $ & $ 0.03 $  & $ 0.072  $ & $
0.064  $  & $ 75 $ & $ 113 $  & $ 0.93 $ & $ 0.91 $ & $ 1.08
$ & $ 0.77 $ \\
\hline

\bf R12 & $ 512 $ & $1.80$ & $ 0.313 $   & $ 1.024  $   & $0.052  $  &
$ 0.29\times10^{-3}    $    & $ 0.235\times10^{-3}  $  & $
9\times 10^{-4} $  & $ 0.03 $ & $ 0.03 $  & $ 0.070  $ & $
0.066  $  & $ 99 $ & $ 116 $  & $ 0.85 $ & $ 0.84 $ & $ 0.77
$ & $ 0.80 $ \\
\hline

\bf R13 & $ 512 $& $1.90$  & $ 0.420 $   & $ 0.98  $   & $0.04  $  & $
0.22\times10^{-3}    $    & $ 0.28\times10^{-3}  $  & $
9\times 10^{-4} $  & $ 0.03 $ & $ 0.03 $  & $ 0.067  $ & $
0.070 $  & $ 124 $ & $ 103 $  & $ 0.83 $ & $ 0.84$ & $ 0.78 $
& $ 0.90 $ \\
\hline

\bf R14 & $ 512 $& $2.10$  & $ 0.741 $   & $ 1.298  $   & $ -0.065  $
& $ 0.16\times10^{-3}    $    & $ 0.42\times10^{-3}  $  & $
9\times 10^{-4} $  & $ 0.03 $ & $ 0.03 $  & $ 0.059  $ & $
0.073 $  & $ 150 $ & $ 70 $  & $ 0.80 $ & $ 0.82 $ & $ 0.61 $
& $ 1.11 $ \\
\hline

\end{tabular}
\caption{Parameters for our DNS runs R1-R14: ${N_{\rm c}}^3$ is the number of
collocation points; $T$ is the temperature; $\rho_{\rm n}/\rho$ is the fraction of
the normal component; $B$ and $B'$ are the  coefficients of the mutual 
friction; $\nu_{\rm n} (\nu_{\rm s})$ is the viscosity of the normal fluid 
(superfluid); $dt$ is the time step; $f_{\rm n} (f_{\rm s})$ is the fixed 
injected energy in the first two shells of the normal fluid (superfluid);
$\lambda_{\rm n} (\lambda_{\rm s})$ is the 
Taylor microscale of the normal-fluid (superfluid); $ Re_{\lambda}^{\rm n}
(Re_{\lambda}^{\rm s})$ is the Taylor-microscale Reynolds numbers for the 
normal fluid (superfluid); $T_{\rm {eddy}}^{\rm n} (T_{\rm {eddy}}^{\rm s})$
is the eddy-turn-over time for the normal fluid (superfluid); and $ \eta_{\rm n}
(\eta_{\rm s})$ is the dissipation length scale for the normal fluid
(superfluid); $k_{\rm {max}}$ is the maximal allowed magnitude of the wave
vectors after the dealiasing. We force the majority component: in the runs
$R1-R6$ we force the superfluid component; in the runs $R7-R8$ we force the
normal-fluid component; in the runs $R9-R14$ we force both the fluids and use the temperature-dependent values for
$\nu_n$ and $\nu_s$ that are given in columns $6$ and $7$, respectively~\cite{viscosity}.}
\label{tab:parameters}
\end{table*}

\subsection{Numerical Results}\label{sec:dns}

We have noted above that, if the normal-fluid and superfluid velocities are
\textit{completely aligned}, the mutual-friction terms do not appear in
Eqs.~(\ref{eq:KH1})-(\ref{eq:KH4}). It is important, therefore, to characterize
the degree of alignment between these velocities. We follow the 2D-HVBK
turbulence study of Ref.~\cite{shukla2015homogeneous}, define the ratio of the
magnitudes of normal-fluid and superfluid velocities $\gamma = \frac{u_{\rm n}}
{u_{\rm s}}$, and then we obtain the probability distribution function
(PDF) $P(\gamma)$ or the cumulative probability distribution function (CPDF)
$Q(\gamma)$. We also obtain the PDF $P(\theta)$, where  $\theta=\cos^{-1}
(\frac{{\bf u}_{\rm n}\cdot {\bf u}_{\rm s}}{u_{\rm n} u_{\rm s}})$ is the
angle between  ${\bf u}_{\rm n}$ and ${\bf u}_{\rm s}$. 

\begin{figure*} [t]
\resizebox{\linewidth}{!}{
\includegraphics[scale=0.60]{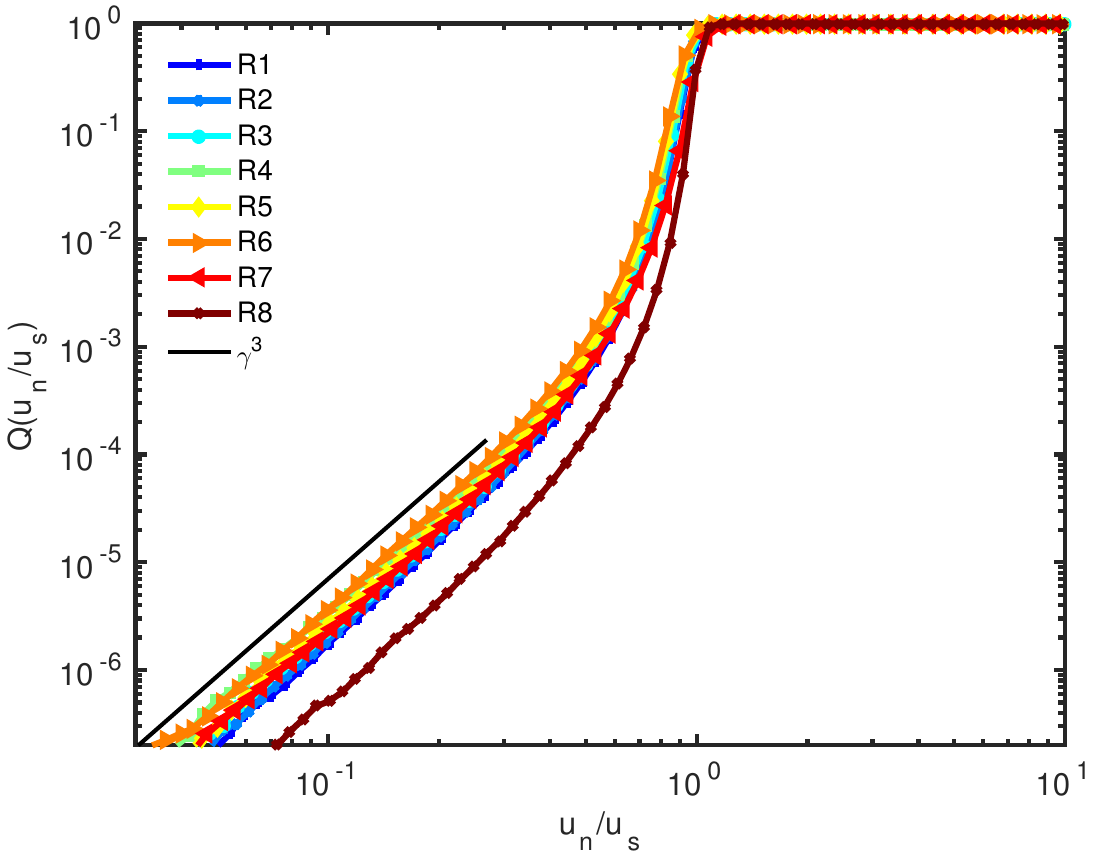}\put(-40,200){\bf(a)}
\includegraphics[scale=0.60]{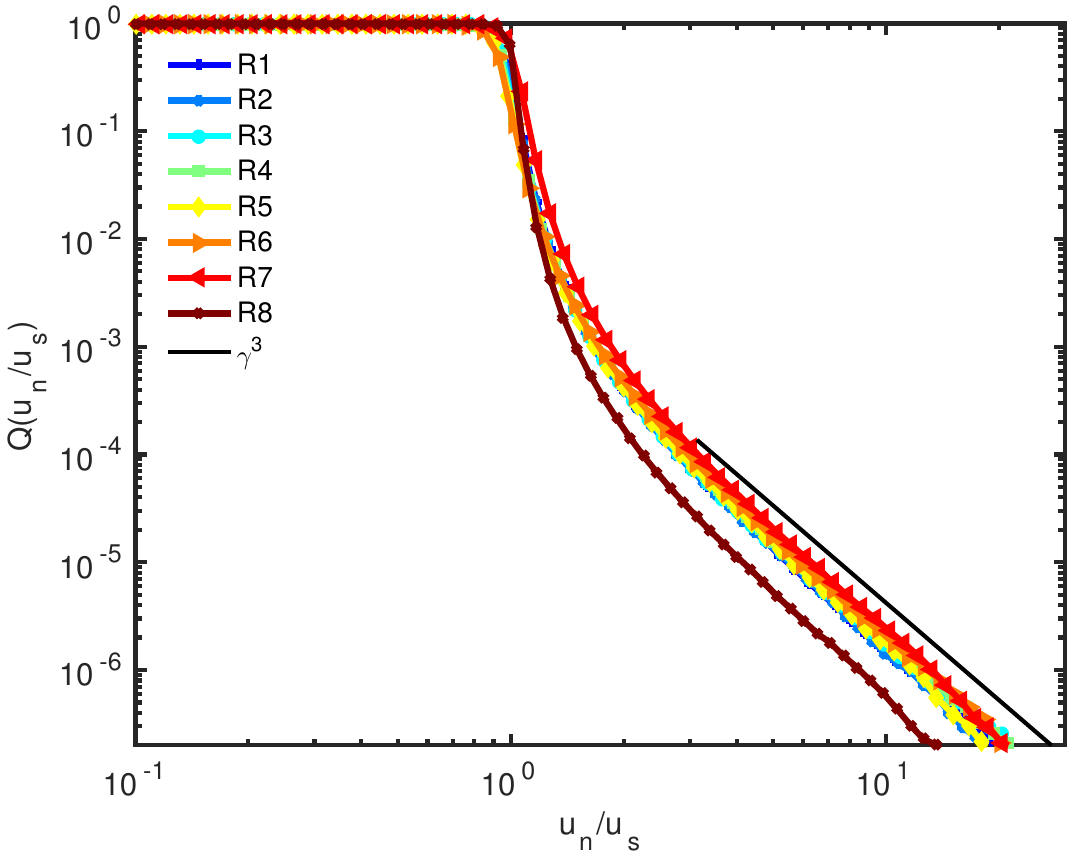}\put(-40,200){\bf(b)}
\includegraphics[scale=0.60]{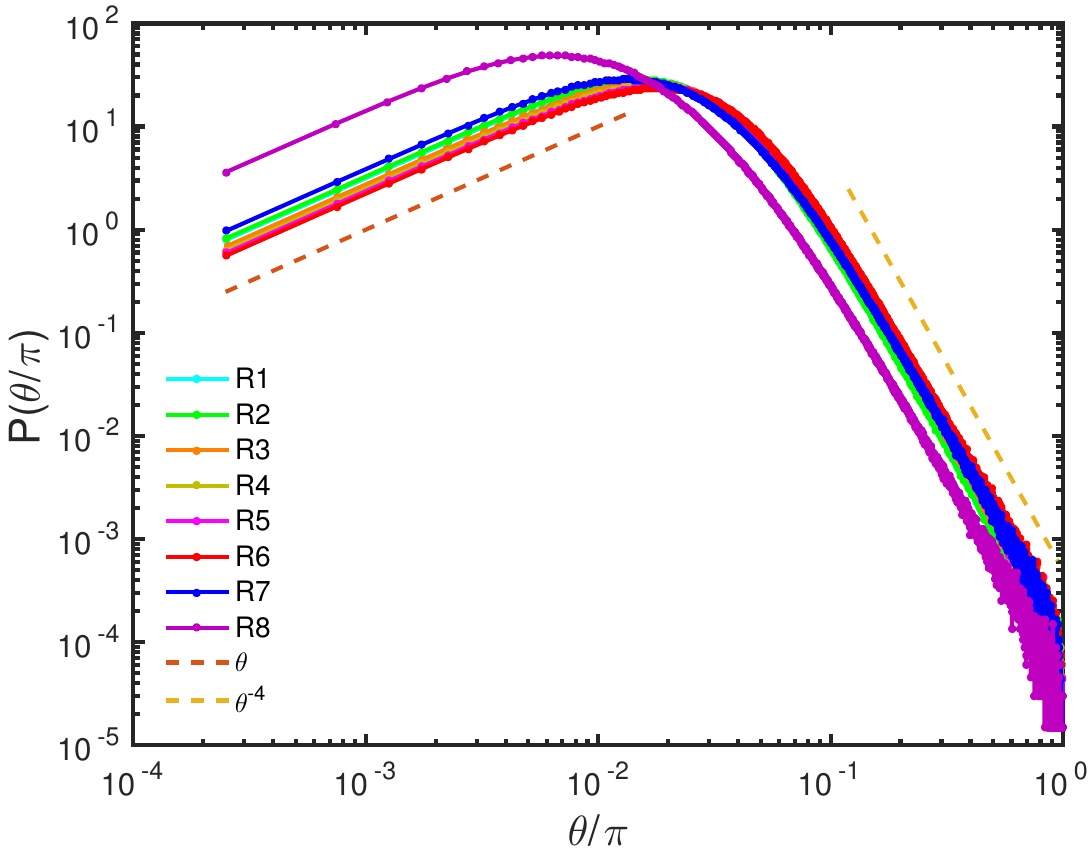}\put(-55,50){\llap{\includegraphics[height=4cm]{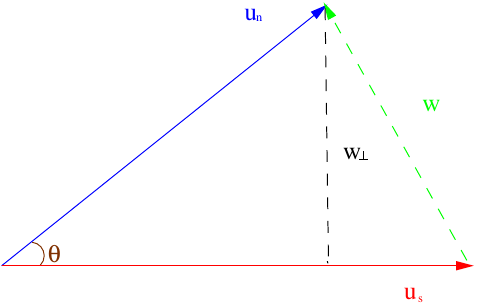}}}\put(-40,200){\bf(c)}}
\resizebox{\linewidth}{!}{
\includegraphics[scale=0.60]{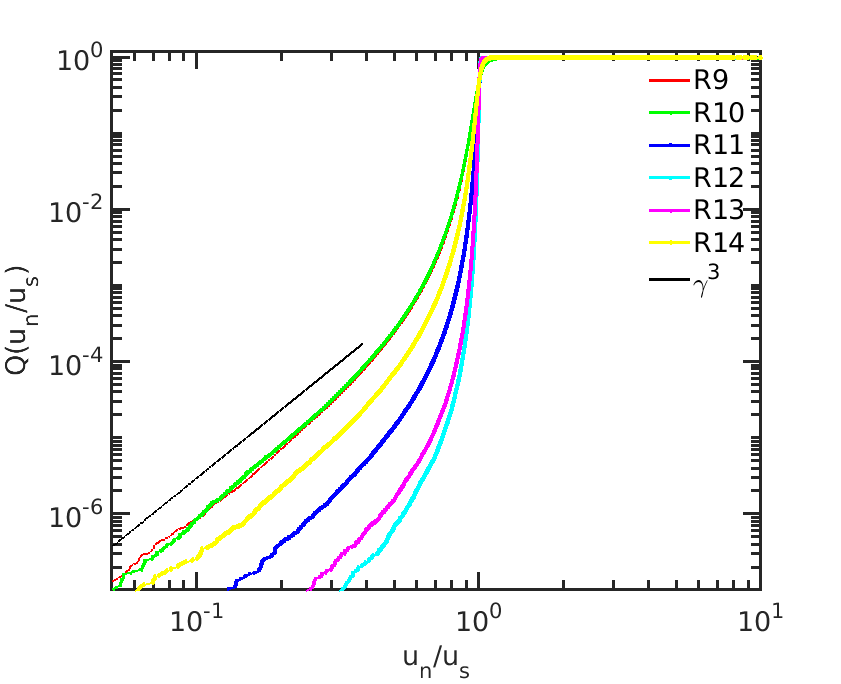}\put(-300,250){\bf(d)}
\includegraphics[scale=0.60]{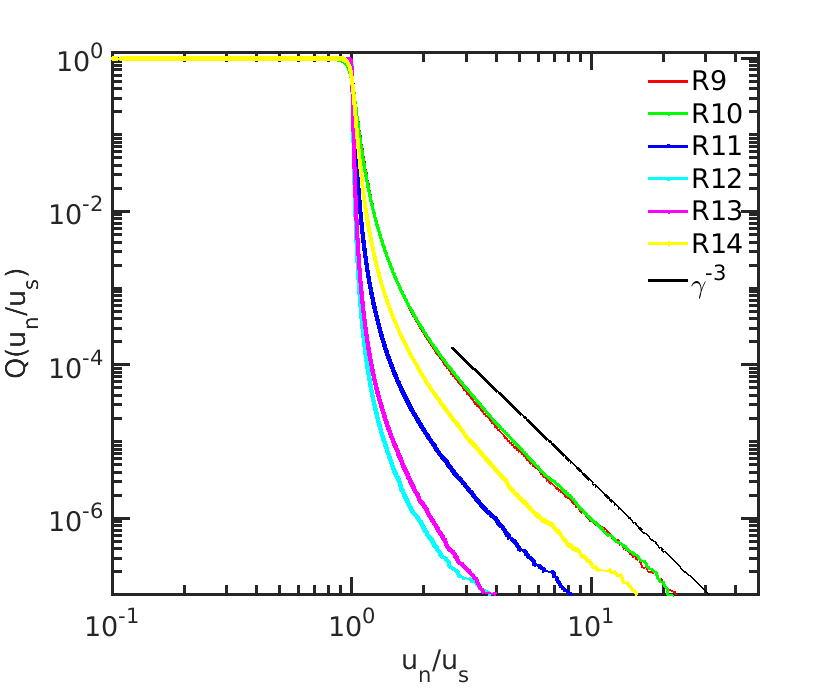}\put(-300,250){\bf(e)}
\includegraphics[scale=0.60]{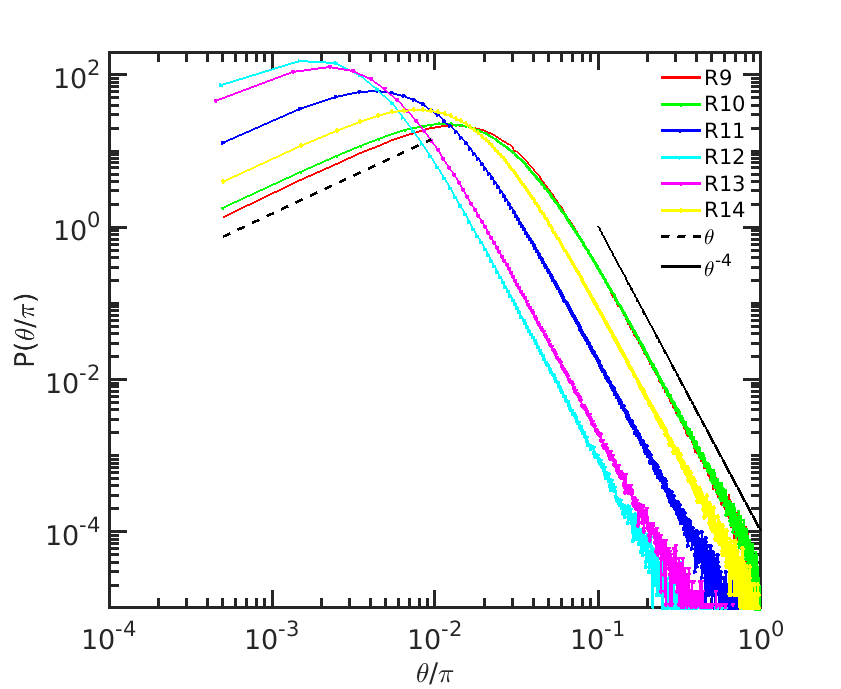}\put(-300,250){\bf(f)}}

\caption{(Color online) Log-Log plots (for runs $R1-R8$) of (a) the complementary cumulative
probability distribution function $(Q(\gamma))$ of $(\gamma) =
\frac{u_n}{u_s}$, the ratio of the normal fluid speed and the
superfluid speed, for $\gamma \ll 1$, where $Q(\gamma) \sim \gamma^3$,
(b) and the CPDF $(Q(\gamma))$, for $\gamma \gg 1$, where $Q(\gamma)
\sim \gamma^{-3}$, and (c) and the probability distribution function
(PDF) $P(\theta)$ of the angle $\theta$ between the normal-fluid
velocity and the superfluid velocity; for $\theta \ll \theta_*,\,
P(\theta)\sim \theta$ and for $\theta_* \ll \theta \ll 1, \, P(\theta)
\sim \theta^{-4}$; (d), (e), and (f) are similar to (a), (b), and (c), 
 respectively (for the runs $R9-R14$ with temperature-dependent viscosities (Table~\ref{tab:parameters}).}

\label{fig:CPDFs}
\end{figure*}

We first present data from our DNS studies of 3D HVBK superfluid turbulence, for the runs
$R1-R8$ (parameters in Table~\ref{tab:parameters}).  In Figs.~\ref{fig:CPDFs} (a)
and (b) we give log-log plots of the CPDF $Q(\gamma)$ versus $\gamma$ for (a)
$\gamma \ll 1$ and (b) $\gamma \gg 1$, respectively. These plots show the
following power-law tails (extending for about a decade given the resolution
of our study) that are consistent with  $Q(\gamma)\sim \gamma^3,
(P(\gamma)\sim \gamma^2),$ for $\gamma \ll 1$ and $Q(\gamma)\sim \gamma^{-3},
(P(\gamma)\sim \gamma^{-4})$ for $\gamma \gg 1$.  Similar results for 2D-HVBK
turbulence (subscript $2D$) have been obtained in
Ref.~\cite{shukla2015homogeneous}: $Q_{2D}(\gamma)\sim \gamma^2,
(P_{2D}(\gamma)\sim \gamma^1),$ for $\gamma \ll 1$ and $Q_{2D}(\gamma)\sim
\gamma^{-2}, (P_{2D}(\gamma)\sim \gamma^{-3})$ for $\gamma \gg 1$.
These exponents appear to be universal, insofar as they do not depend on the 
parameters (like  $B$ and $B'$) in 3D- and 2D-HVBK superfluid turbulence; however,
these exponents depend on the dimension $d$.

In Fig.~\ref{fig:CPDFs} (c) we display log-log plots of the PDF $P(\theta)$ for
all our DNS runs $R1-R8$ (Table~\ref{tab:parameters}). These show that
$P(\theta) \sim \theta$, for $\theta \ll 1$ and $\theta \ll \theta^*$; and
$P(\theta) \sim \theta^{-4}$, for $\theta \ll 1$ and $\theta \gg \theta^*$
(given the resolution of our study, these scaling forms extend for slightly
more than a decade in $\theta$). Furthermore, these power-law exponents do not
depend on parameters such as $B$ and $B'$ and are, in this sense, universal. Figures~\ref{fig:CPDFs} (d), (e), and (f) show the CPDF $Q(\gamma)$ and the PDF $P(\theta)$, respectively, for the runs $R9-R14$ with temperature-dependent viscosities (Table~\ref{tab:parameters}); these are similar to Figs.~\ref{fig:CPDFs} (a), (b), and (c). The exponents for the asymptotic behaviors of the CPDFS and PDFs in Figs. \ref{fig:CPDFs} (d), (e) and (f) are the same as those of their counterparts in Figs. \ref{fig:CPDFs} (a), (b), and (c) respectively, with some minor changes in the tails, which arise because of the differences in $\nu_n/\nu_s$ for the runs $R9-R14$ with temperature-dependent viscosities (Table~\ref{tab:parameters}).

We now show that the power-law regimes (and the exponents that characterize
them) in the plots of Fig.~\ref{fig:CPDFs} can be obtained by making
reasonable assumptions about the joint probability distribution function (JPDF)
$\mathcal{P}(u_{\rm n},u_{\rm s})$, from which we can obtain $P(\gamma)$ as
follows:
\begin{equation}
P(\gamma)=\int_0^\infty \int_0^\infty du_{\rm n} du_{\rm s} \delta(\gamma - 
\frac{u_{\rm n}}{u_{\rm s}}) \mathcal{P}(u_{\rm n},u_{\rm s}) .
\label{eq:PGammadef}
\end{equation}

For $\gamma \gg 1$ and $\gamma \ll 1$, one or the other fluid dominates, so we
expect that the normal-fluid and superfluid velocities should be nearly
uncorrelated (this is not true if $\gamma \simeq 1 $). Therefore, we can make
the approximation $\mathcal{P}(u_{\rm n},u_{\rm s}) \sim P(u_{\rm n}) P(u_{\rm
s})$ (we have checked this numerically), for
%
%
$\gamma \gg 1$ and $\gamma \ll 1$ [$P(u_{\rm n})$ and $P(u_{\rm s})$ are the 
PDFs of $u_{\rm n}$ and $u_{\rm s}$, respectively], that yields 
\begin{equation}
P(\gamma) \sim \int_0^\infty \int_0^\infty du_{\rm n} du_{\rm s} \delta(\gamma-
\frac{u_{\rm n}}{u_{\rm s}}) P(u_{\rm n})P(u_{\rm s}).
\label{eq:PGamma1}
\end{equation}
We find that the components of the normal and superfluid velocities have
PDFs that are very close to Gaussian ones in HVBK superfluid turbulence, like the PDFs of
components of the fluid velocity in classical-fluid turbulence (see, e.g., 
Refs.~\cite{pandit2017overview,dhar1997some} and references therein); therefore, in $d$ 
spatial dimensions, the magnitudes of these velocities should have the
Maxwellian PDFs $P(u_{\rm n})\sim C_{\rm n}u_n^{d-1} \exp(-\frac{u_{\rm n}^2}
{\sigma_{\rm n}^2})$ and $P(u_{\rm s})\sim C_{\rm s}u_{\rm s}^{d-1} 
\exp(-\frac{u_{\rm s}^2}{\sigma_{\rm s}^2})$, where $C_{\rm n} (C_{\rm s})$ 
and $\sigma_{\rm n} (\sigma_{\rm s})$ are, respectively, the normalization constant and 
standard deviation for the velocity of the normal fluid (superfluid). If we
substitute these Maxwellian forms in Eq.~(\ref{eq:PGamma1}) and integrate over
$u_{\rm n}$ and $u_{\rm s}$ we get 
\begin{equation}
P(\gamma)=\frac{C_n C_s}{2}\frac{\gamma^{d-1}}{(\frac{\gamma^2}{\sigma_n^2}+
\frac{1}{\sigma_s^2})^d}\Gamma(d),  
\label{eq:PGamma2}
\end{equation}
whence we obtain $P(\gamma)\sim \gamma^{d-1}$, for $\gamma \ll1$, and $P(\gamma)
\sim \gamma^{-d-1}$, for $\gamma \gg 1$; these exponents are consistent 
with the results we have given above, for 3D-HVBK superfluid turbulence, and the results
presented in Ref.~\cite{shukla2015homogeneous}, for 2D-HVBK superfluid turbulence.

To obtain the scaling forms of the PDF $P(\theta)$ at small and large values of
$\theta$ (Fig.~\ref{fig:CPDFs} (c)) we note that $\sin{\theta}=
\frac{w_{\perp}}{u_{\rm n}}$ (inset of Fig.~\ref{fig:CPDFs} (c)), where ${\bf
w}={\bf u}_{\rm n}-{\bf u}_{\rm s}$ and $w_{\perp}=u_{{\rm n}_{\perp}}$.  For
$\theta \ll 1$, $\sin{\theta}\sim \theta$ and $u_{{\rm n}_{\perp}} = a_{{\rm
n}_{\perp}}t_{\rm n} $; here, $t_{\rm n} \ll 1$ and $a_{{\rm n}_{\perp}}$ is
the normal component of the acceleration of the normal fluid.  Clearly, 
\begin{equation}
P(\theta)=\int\int du_{\rm n} da_{{\rm n}_{\perp}}\delta(\theta-\frac{a_{{\rm n}_{\perp}t_{\rm n}}}{u_{\rm n}}) \mathcal{P}(u_{\rm n},a_{{\rm n}_{\perp}}),
\label{eq:Ptheta}
\end{equation}
where $\mathcal{P}(u_{\rm n},a_{{\rm n}_{\perp}})$ is  the joint PDF of 
$u_{\rm n}$ and $a_{{\rm n}_{\perp}}$. We now make the approximation 
\begin{equation}
\mathcal{P}(u_{\rm n},a_{{\rm n}_{\perp}}) \sim P(u_{\rm n}) P(a_{{\rm n}_{\perp}}), 
\label{eq:decouple}
\end{equation}
which can be justified within the framework of the Kolmogorov theory of 1941
(K41)~\cite{frisch1995turbulence} as follows (our arguments follow those in
Ref.~\cite{bhatnagar2016deviation}, which obtains the PDF of the angle between
the  Eulerian velocity of a turbulent fluid and the velocity of an inertial
particle that is advected by this fluid): K41 assumes that, in a homogeneous,
isotropic, and statistically steady turbulent flow, the only large-length-scale
property that is of importance at small length scales is $\epsilon$, the rate
of energy dissipation. Viscous dissipation becomes significant at length scales
smaller than the K41 dissipation scale $\eta_d = [\nu^3/\epsilon]^{1/4}$; at
such scales the typical fluid acceleration is $a_* =\epsilon^{3/4}\nu^{-1/4}$,
whereas the dissipation-scale velocity $u_{\eta_d}=(\epsilon\nu)^{1/4}$.  In
the large-Reynolds-number limit, i.e., $\nu \to 0$, in a 3D turbulent fluid,
$\epsilon$ goes to a positive constant (the dissipative anomaly); therefore,
$a_*$  is much larger than typical accelerations because of large-scale fluid
motion; by contrast, $u_{\eta_d}$ is much smaller than large-scale velocities.
In summary, the normal component of the fluid acceleration can be large at
small scales, where it is determined, principally, by small-scale properties of
the flow; in contrast, dominant fluid velocities are determined by large-scale
motions. The separation of length scales in the K41 theory then suggests that,
to a good approximation, $a_*$ and $u_{\eta_d}$ are statistically
independent, so their JPDF can be approximated by the product of their
respective PDFs. This argument can be applied, \textit{mutatis mutandis}, to
the normal fluid in 3D HVBK turbulence to justify Eq.~(\ref{eq:decouple}). 

We have noted above that, in the HVBK model, $P(u_{\rm n})$ is very well
approximated by the Maxwellian distribution $P(u_{\rm n}) = C_{\rm n}
u_{\rm n}^{d-1} \exp(\frac{-u_{\rm n}^2}{2\sigma^2})$; our numerical data are 
consistent with $P(a_{{\rm n}_{\perp}}) = B_1 a_{{\rm n}_{\perp}}^{d-2} 
\exp({-B_2 a_{{\rm n}_{\perp}}^2})$, where $C_{\rm n}, B_1$, and $B_2$
are constants (this PDF has a similar form in classical-fluid turbulence
~\cite{bhatnagar2016deviation}). If we use these forms for $P(u_{\rm n})$
and $P(a_{{\rm n}_{\perp}})$, along with Eqs.~(\ref{eq:Ptheta}) and
(\ref{eq:decouple}), and then integrate over $u_{\rm n}$, we get
\begin{equation}
P(\theta)= \int da_{{\rm n}_{\perp}} t_{\rm n}^d C_{\rm n} B_1 \frac{a_{{\rm n}_{\perp}}^{d+1}} {\theta^{d+1}}\exp({-B_2 a_{{\rm n}_{\perp}}^2}) \exp({\frac{-a_{{\rm n}_{\perp}}^2 t_{\rm n}^2}{2\theta^2\sigma^2}}).
\label{eq:Pthetasemifinal}
\end{equation}
If we define the angular scale $\theta_* = \frac{a_* t_{\rm n}}{{\sqrt 2}\sigma}$
and the dimensionless variables $X = \frac{\theta}{\theta_*}$ and $Y =
\frac{a_{{\rm n}_{\perp}}}{a_*}$, then Eq.~(\ref{eq:Pthetasemifinal}) becomes
\begin{eqnarray}
P(\theta)&=&\int dY t_{\rm n}^d C_{\rm n} B_1 \frac{Y^{d+1}}{\theta^{d+1}}
a_*^{d+2} \exp(-B_2 Y^2 a_*^2) \nonumber \\
&\times&\exp(\frac{-Y^2}{X^2}) .
\label{eq:Pthetafinal}
\end{eqnarray}

We now consider the ranges (a) $0 \leq \theta \ll \theta_*$, $X \ll 1$ and (b)
$\theta_* \ll \theta \ll 1$, $1 \ll X$. Case (a): the leading term of
Eq.~(\ref{eq:Pthetafinal}) is $P(\theta)\sim \int_0^X dY t_n^d C_n B_1
\frac{Y^{d+1}}{\theta^{d+1}}$, which can be simplified to get $P(\theta)\sim
\theta^{d/3}$, i.e., $P(\theta) \sim \theta$ in $d=3$ for $0 \ll \theta \ll
\theta_*$. Case (b): In this range $\exp{\frac{-Y^2}{X^2}} \approx 1$ so
Eq.~(\ref{eq:Pthetafinal}) yields $P(\theta)= \theta^{-(d+1)}\int dY 
t_{\rm n}^d C_{\rm n} B_1 \frac{Y^{d+1}}{a_*}^{d+2} \exp(-B_2 Y^2 a_*^2)$,
whence we get $P(\theta)\sim \theta^{-(d+1)}$, i.e., $P(\theta)\sim 
\theta^{-4}$ in $d=3$, in the range $\theta_* \ll \theta \ll 1$. The power
laws in the ranges (a) and (b) are consistent with our numerical results in
Fig.~\ref{fig:CPDFs}.

 \begin{figure*} [t]
 \resizebox{\linewidth}{!}{
 \includegraphics[scale=0.7]{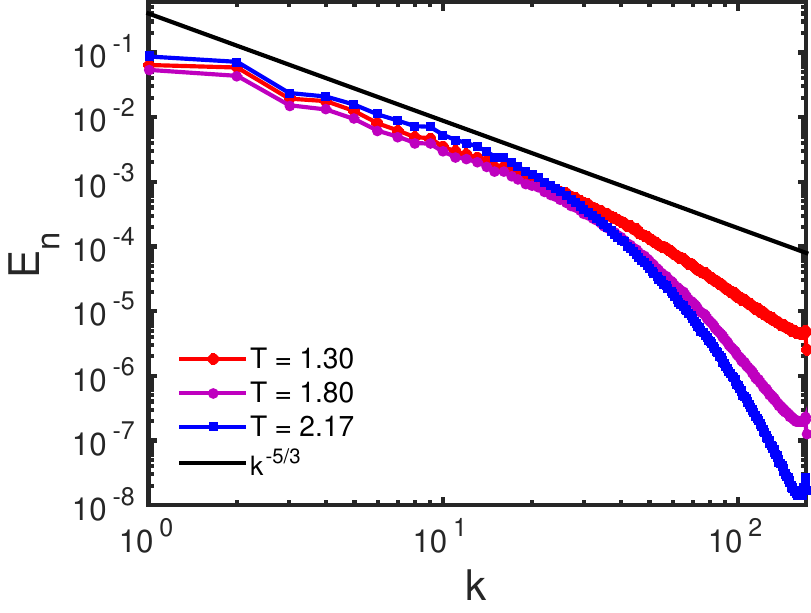}\put(-70,190){\bf(a)} 
 \includegraphics[scale=0.7]{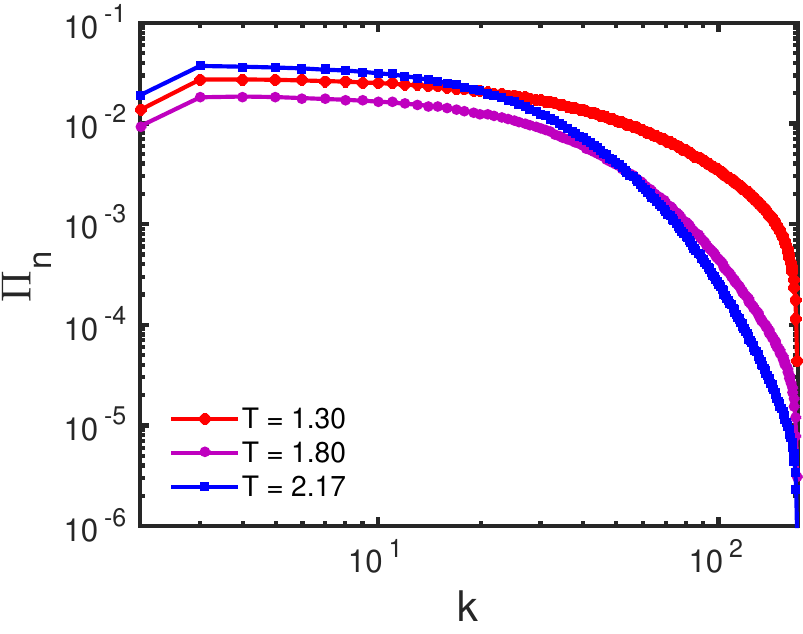}\put(-100,190){\bf(c)}
 \includegraphics[scale=0.7]{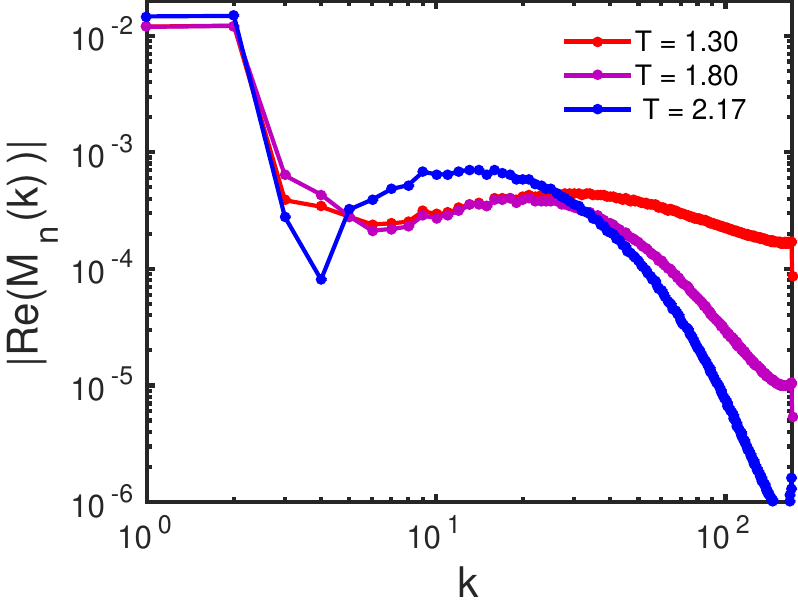} \put(-100,185){\bf(e)}}

 \resizebox{\linewidth}{!}{
 \includegraphics[scale=0.7]{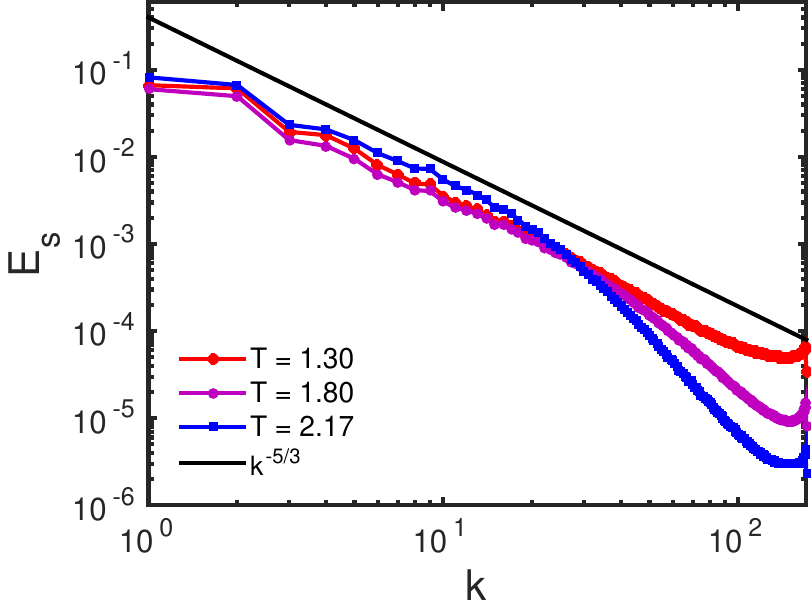}\put(-70,190){\bf(b)}
 \includegraphics[scale=0.7]{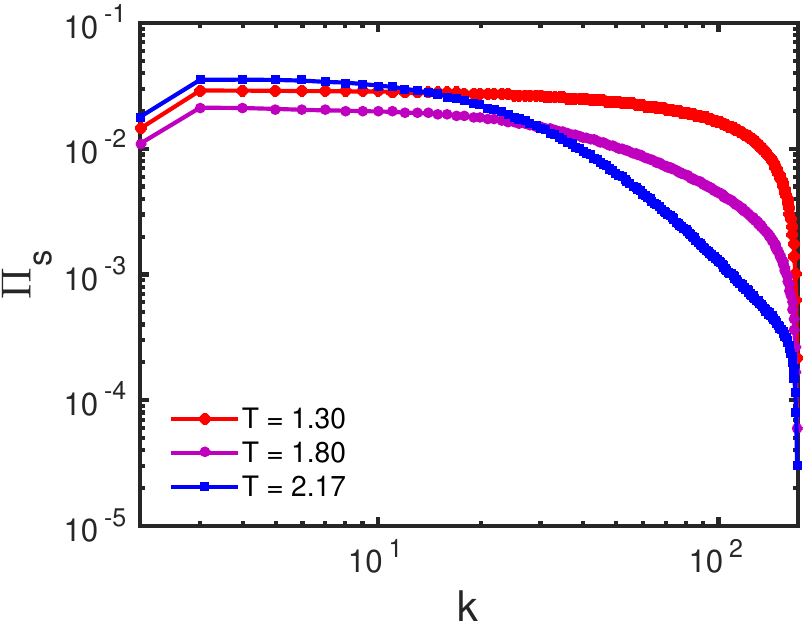}\put(-100,190){\bf(d)}
 \includegraphics[scale=0.7]{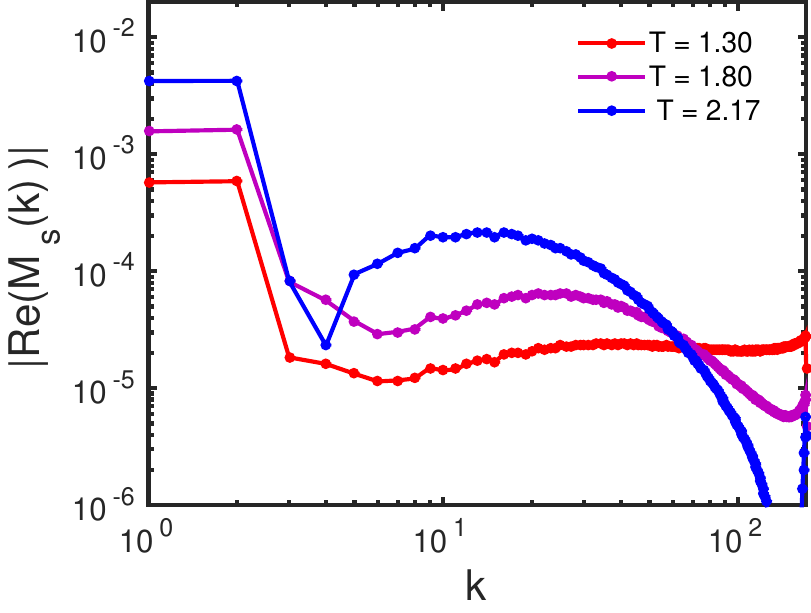} \put(-100,185){\bf(f)}}

 \resizebox{\linewidth}{!}{
 \includegraphics[scale=0.5]{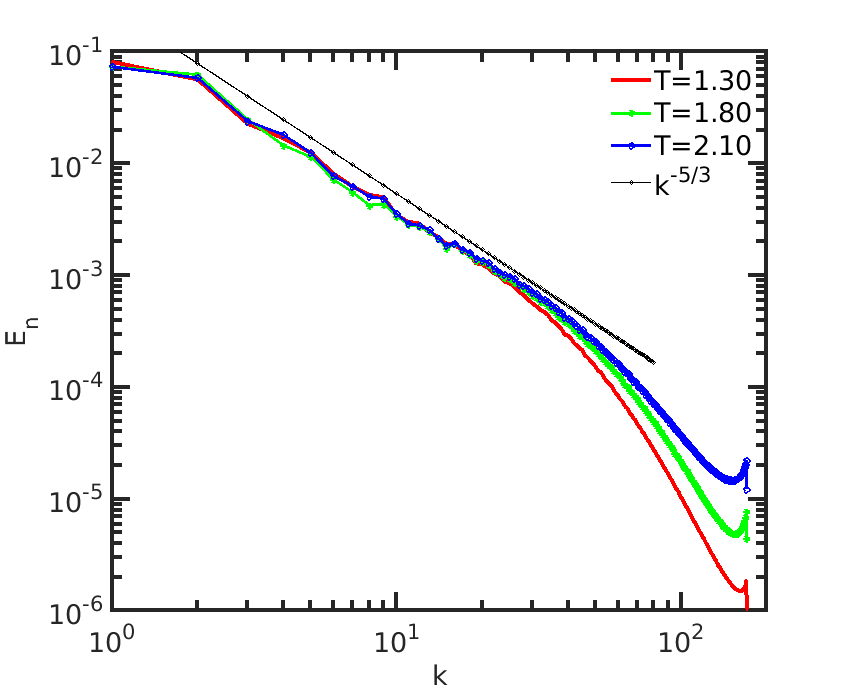}\put(-250,200){\bf(g)} 
 \includegraphics[scale=0.5]{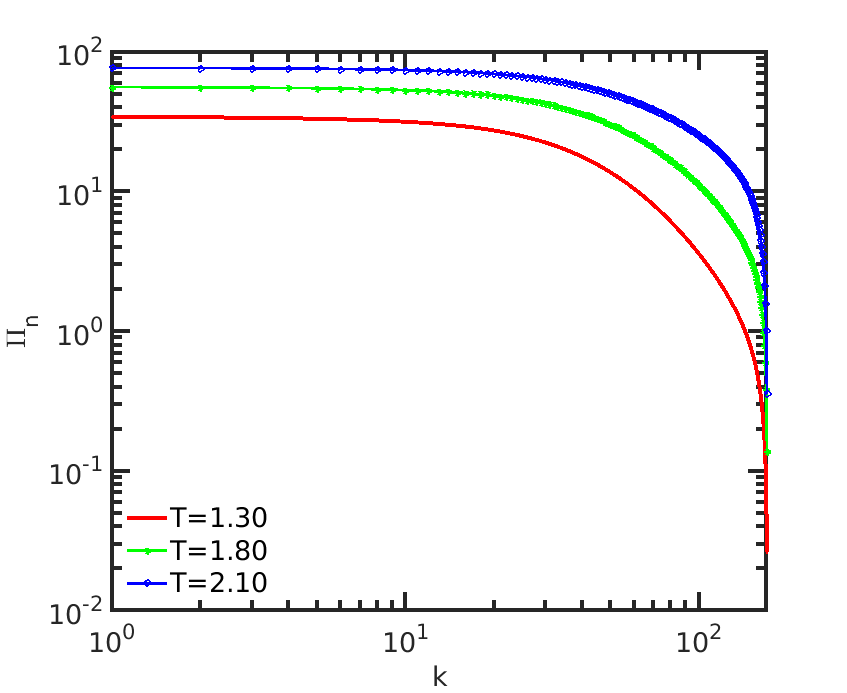}\put(-250,200){\bf(h)}
 \includegraphics[scale=0.5]{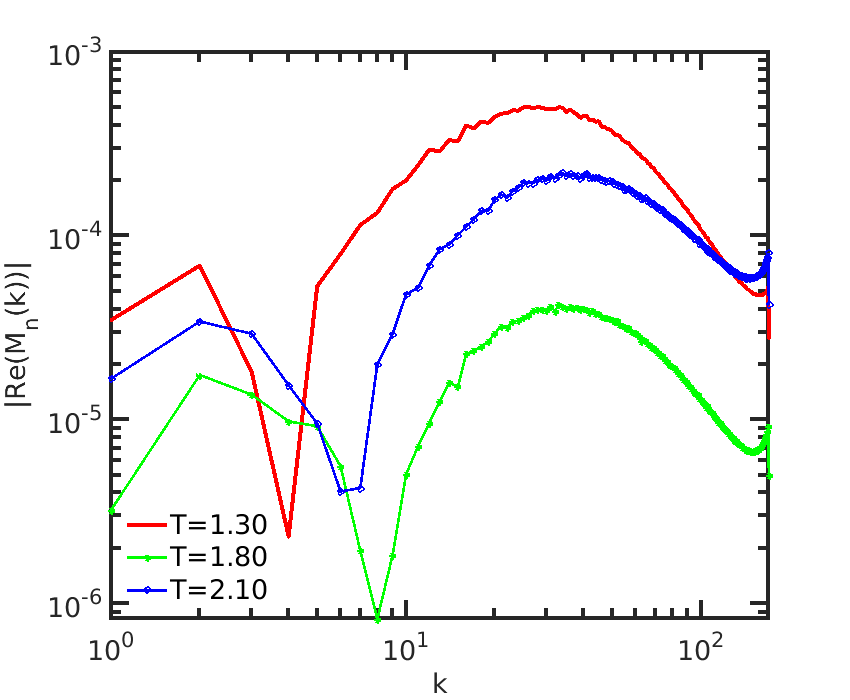} \put(-250,200){\bf(i)}}

\caption{(Color online) Log-Log plots (for $T=1.30$, red lines, $T=1.80$ 
magenta lines,
 and $T=2.17$ blue lines) versus the wavenumber $k$ of  (a) the
 energy spectrum $E_{\rm n}(k)$, for the normal fluid, (b) the energy
 spectrum $E_{\rm s}(k)$, for the superfluid [the black lines indicate
 the Kolmogorov 1941 (K41) scaling form $\sim k^{-5/3}$], (c) the
 energy flux $(\Pi_{\rm n})(k)$, for the normal fluid, (d) the energy
 flux $(\Pi_{\rm s})(k)$, for the superfluid, (e) the absolute value of
 the real component of the mutual-friction transfer
 $(\left|Re(\rho_{\rm s} {\bf f}_{mf}  \cdot {\bf u}_{\rm n})\right|)$,
 for the normal fluid, and $(f)$ the absolute value of the real
 component of the mutual-friction transfer  $(\left|Re(\rho_{\rm n}
 {\bf f}_{mf} \cdot {\bf u}_{\rm s}) \right|)$, for the superfluid. (g), (h), and (i) are similar to (a), (c), and (e), respectively, but for the runs $R9, R12$, and $R14$ with temperature-dependent viscosities (Table~\ref{tab:parameters}).}

\label{fig:spectra}
 \end{figure*}

 \begin{figure*} [t]
 \resizebox{\linewidth}{!}{
 \includegraphics[scale=0.67]{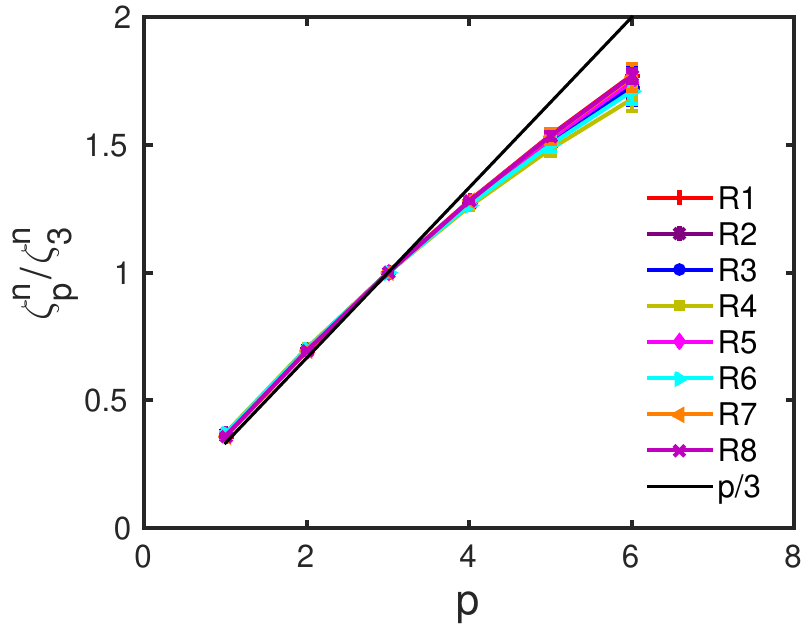}\put(-70,150){\bf(a)}
\includegraphics[scale=0.5]{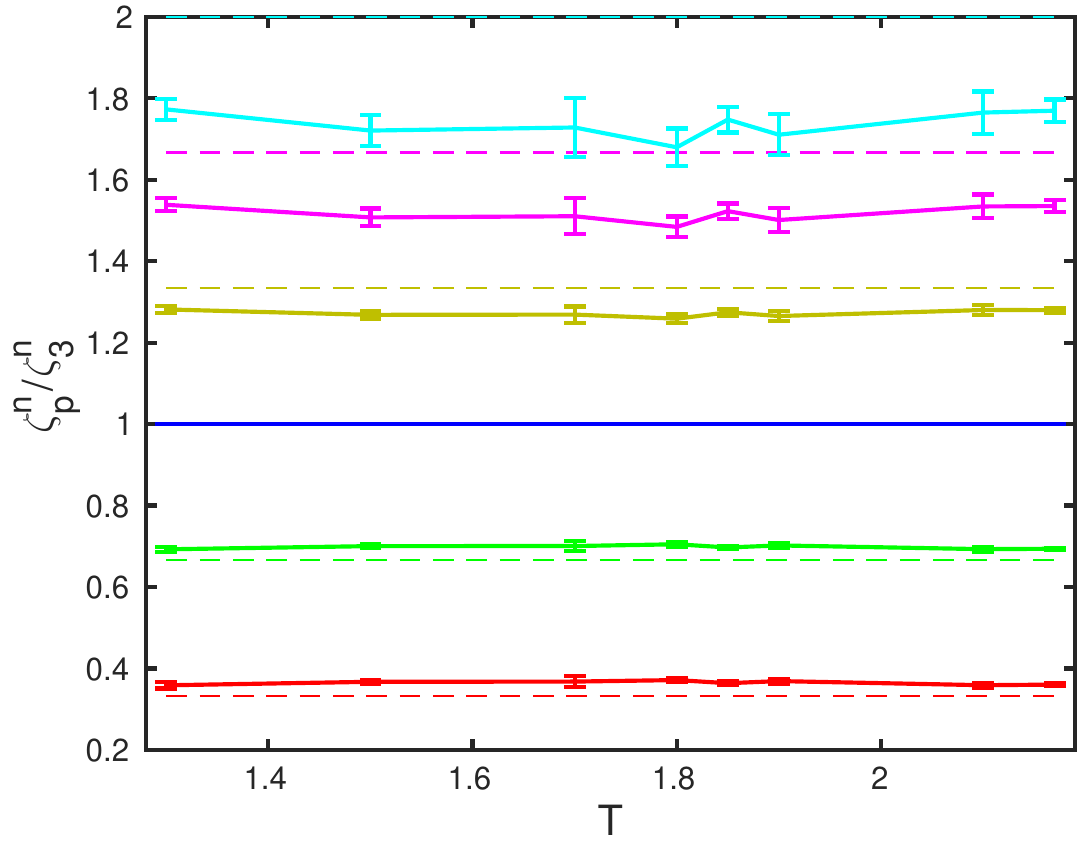}\put(-100,180){\bf(c)}
 \includegraphics[scale=0.67]{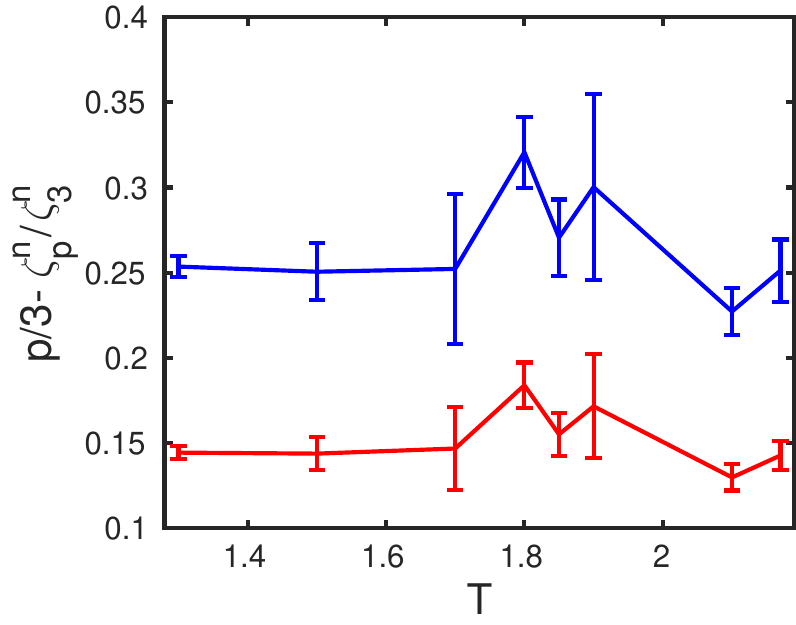} \put(-70,185){\bf(e)}}

 \resizebox{\linewidth}{!}{
 \includegraphics[scale=0.67]{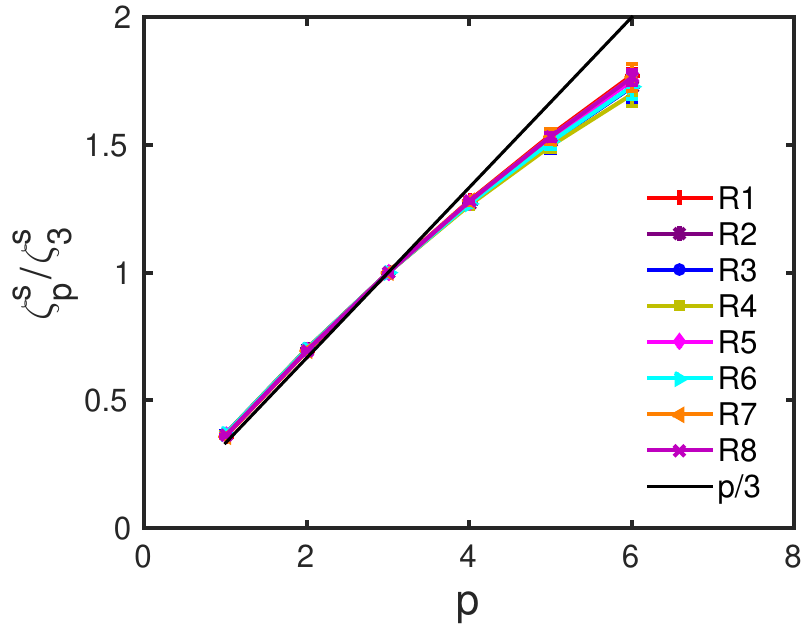}\put(-70,150){\bf(b)}
 \includegraphics[scale=0.5]{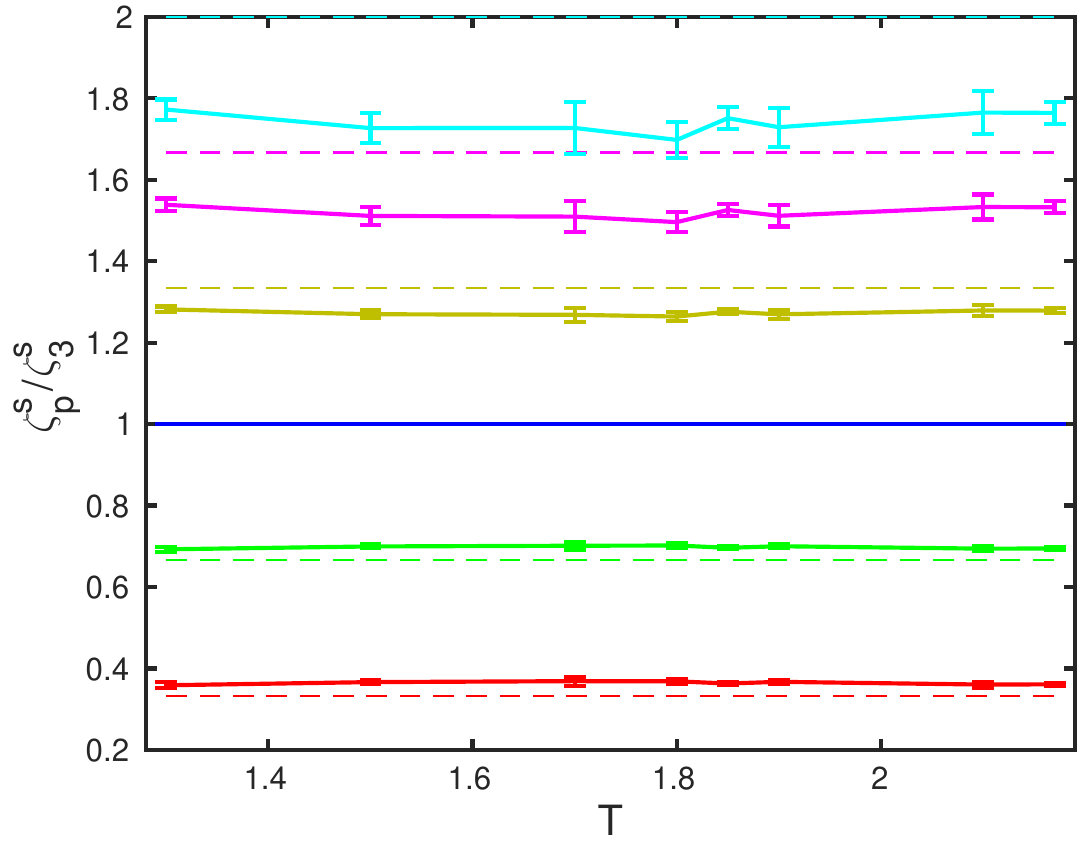}\put(-100,180){\bf(d)}
\includegraphics[scale=0.67]{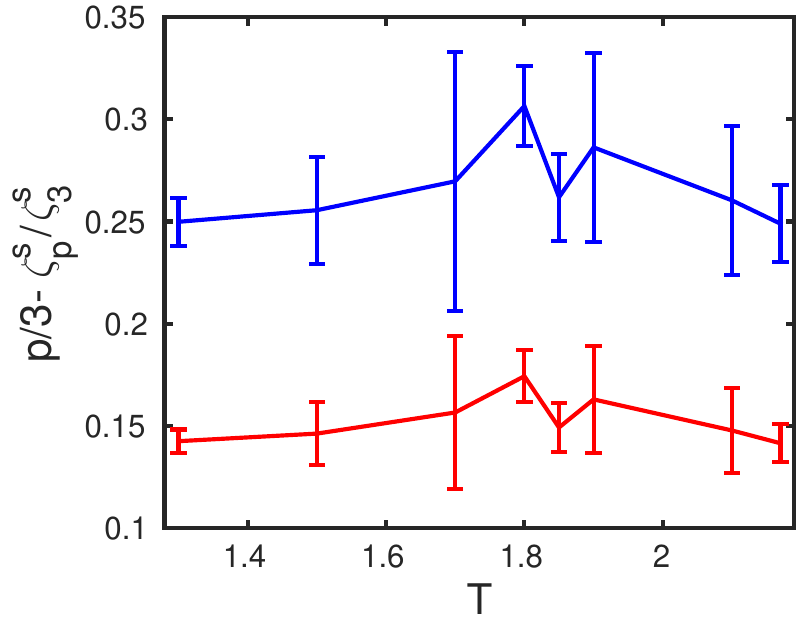} \put(-70,185){\bf(f)}}

\resizebox{\linewidth}{!}{
 \includegraphics[scale=0.6]{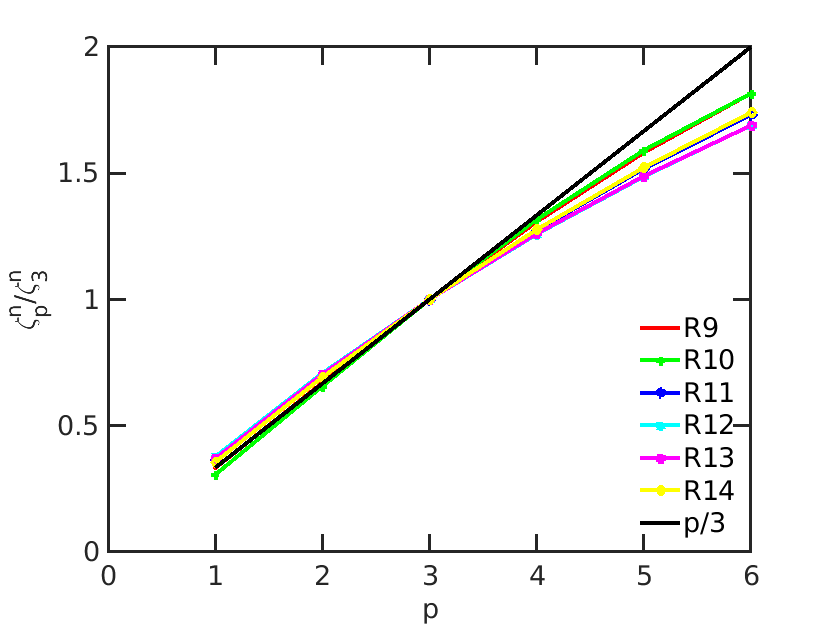}\put(-260,220){\bf(g)}
\includegraphics[scale=0.6]{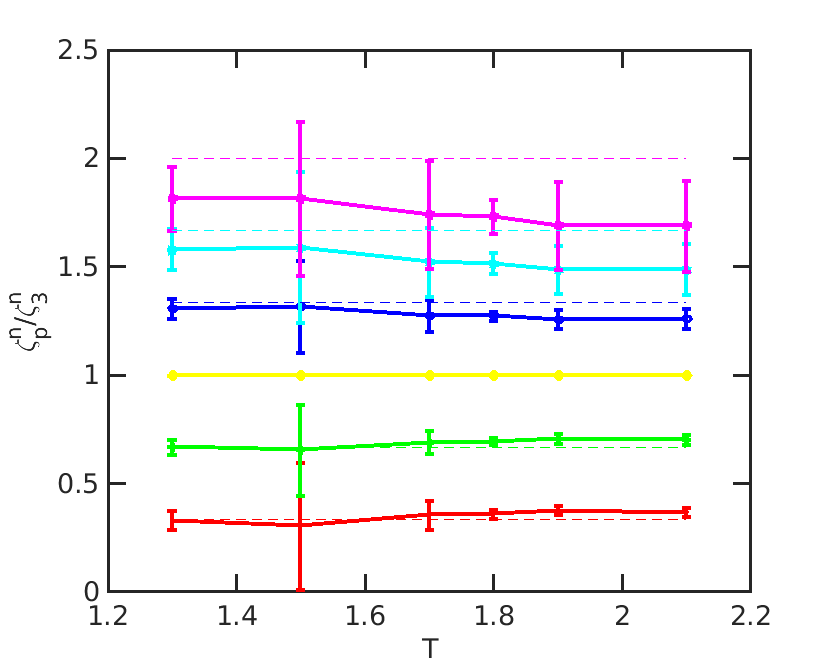}\put(-260,220){\bf(h)}
 \includegraphics[scale=0.6]{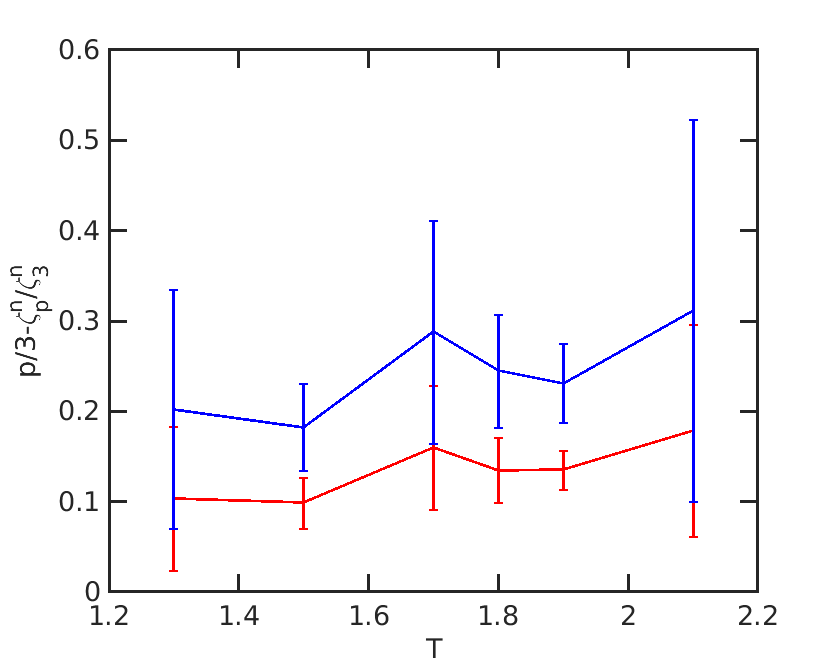} \put(-260,220){\bf(i)}}

 \caption{(Color online) Plots versus the order $p$ of the longitudinal-structure-function
 exponent ratios  (a) for the normal fluid $({\zeta_{p}^{\rm
 n}}/{\zeta_{3}^{\rm n}})$ and (b) for the superfluid $({\zeta_{p}^{\rm
 s}}/{\zeta_{3}^{\rm s}})$ for runs $R1-R8$; $1 \leq p \leq 6$; and we
 use the extended-self-similarity (ESS) method.  Plots of these ratios
 versus the temperature $T$ for (c) the normal fluid and (d) the
 superfluid and $p=1$ (red lines),  $p=2$ (green lines), $p=3$ (blue
 lines),  $p=4$ (yellow lines),  $p=5$ (pink lines), and $p=6$ (cyan
 lines); the dashed lines show the K41 predictions.  Plots versus $T$
 of the intermittency exponents (e) $\mu_p^{\rm n} = p/3-\zeta_{p}^{\rm
 n}/\zeta_{3}^{\rm n}$, for the normal fluid, and (f) $\mu_p^{\rm s} =
 \zeta_{p}^{\rm s}/\zeta_{3}^{\rm s}$ for the superfluid, for $p = 5$
 (red lines) and $p = 6$ (blue lines); (g), (h), and (i) are similar to (a), (c), and (e), respectively, but for the runs $R9-R14$ with temperature-dependent viscosities (Table~\ref{tab:parameters}).}

\label{fig:PiZeta}
 \end{figure*}

In Figs. ~\ref{fig:spectra} (a) and (b) we present log-log plots of the energy spectra 
\begin{eqnarray}
{E}_{\rm n}(k) &=& \sum_{k-1/2 < k' < k + 1/2}
{\bf u}_{\rm n}({\bf k}')\cdot{\bf u}_{\rm n}({-\bf k}')
\nonumber \\
{E}_{\rm s}(k) &=& \sum_{k-1/2 < k' < k + 1/2}
{\bf u}_{\rm s}({\bf k}')\cdot{\bf u}_{\rm s}({-\bf k}')
\label{eq:mftrans}
\end{eqnarray}
for the normal fluid and the superfluid, respectively, for $T=1.30$, $T=1.80$,
and  $T = 2.17$; 
the black lines indicate the Kolmogorov 1941 (K41) scaling form $\sim k^{-5/3}$. Figure~\ref{fig:spectra} (g) shows log-log plots versus $k$ of the energy spectrum $E_n(k)$; this is similar to Fig. ~\ref{fig:spectra} (a), but for the runs with temperature dependent viscosities (Table~\ref{tab:parameters}). Figure~\ref{fig:spectra} (g) shows that the tails of the spectra move up as we increase the temperature; this is similar to the results for these spectra in Ref.~\cite{biferale2018turbulent}. 
In Figs.~\ref{fig:spectra} (c) and (d)  we present log-log plots of the 
energy-flux spectra 
\begin{eqnarray}
\Pi_{\rm n} &=& \bigg \langle \int_{k}^{k_{\text{max}}}\mathcal{T}_{\rm n}(k',t)dk' \bigg \rangle , \nonumber \\ 
\Pi_{\rm s} &=& \bigg \langle \int_{k}^{k_{\text{max}}}\mathcal{T}_{\rm s}(k',t)dk' \bigg \rangle ,
\label{eq:fluxes}
\end{eqnarray}
for the normal fluid and the superfluid, respectively. Fig.~\ref{fig:spectra} (h) is similar to Fig.~\ref{fig:spectra} (c) but for the runs with temperature dependent viscosities (Table~\ref{tab:parameters}); the constant-energy-flux
parts of these plots indicate the extents of the inertial ranges in our DNSs
for $T=1.30$, $T=1.80$, and  $T = 2.17$. Here, $\mathcal {T}_{\rm n}(k',t)$
and $\mathcal {T}_{\rm
s}(k',t)$ are energy-transfer terms in Fourier space because of the triadic
interactions in the normal fluid and superfluid, respectively. The parameters
for these runs are given in Table~\ref{tab:parameters}; we have taken the
dependence of $B, \, B'$, and $\rho_n/\rho_s$ on the temperature $T$ from the
measurements of Ref.~\cite{barenghi1983mfriction} on superfluid $^4$He; 
therefore, our results are applicable to measurements of the statistical properties
of superfluid turbulence in this system.  In Figs.~\ref{fig:spectra} (e) and
(f) we present log-log plots of the absolute values of the real part of the
mutual-friction transfer terms 
\begin{eqnarray}
\mathcal{M}_{\rm n}(k,t) &=& \sum_{k-1/2 < k' < k + 1/2}
\rho_{\rm s}{\bf f}_\text{mf}({\bf k}')\cdot{\bf u}_{\rm n}({-\bf k}') ,
\nonumber \\
\mathcal{M}_{\rm s}(k,t) &=& \sum_{k-1/2 < k' < k + 1/2}
\rho_{\rm n}{\bf f}_\text{mf}({\bf k}')\cdot{\bf u}_{\rm n}({-\bf k}') , 
\label{eq:mftrans}
\end{eqnarray}
for the normal-fluid and superfluid components, respectively.  We observe that,
if we increase the temperature, the mutual-friction transfer for the superfluid
increases. 

The longitudinal velocity structure functions are
\begin{equation}
S_p^{\alpha}(l)=\left \langle \left|\big({\bf u}_{\alpha}({\bf r}+{\bf l})-
{\bf u}_{\alpha}({\bf r})\big)\cdot \bf {\hat{l}} \right |^p \right \rangle ;
\end{equation} 
here, $\alpha =$ n or s, for the normal fluid and superfluid, respectively. In the
inertial range $\eta_d \ll l \ll L$, $S_p^{\alpha}(l) \sim
l^{\zeta_p^{\alpha}}$; we can use this scaling form to extract the exponents
$\zeta_p^{\alpha}$ from $S_p^{\alpha}(l)$. Furthermore, we can extend the
scaling range by using the extended-self-similarity (ESS)
method~\cite{pandit2017overview,dhar1997some,benzi1993extended,chakraborty2010extended}
to calculate the exponent ratio $\zeta_p^{\alpha}/\zeta_3^{\alpha}$ from the
inertial-range slopes of log-log plots of $S_p^{\alpha}(l)$ versus
$S_3^{\alpha}(l)$. In Figs.~\ref{fig:PiZeta} (a) and (b) we plot, respectively,
the exponent ratios $\zeta_p^{\rm n}/\zeta_3^{\rm n}$ and $\zeta_p^{\rm s}/
\zeta_3^{\rm s}$ versus the order $p$ ($p \leq 6$). Figures~\ref{fig:PiZeta}
(c), and (d) show the plots of these  exponent ratios versus the temperature
$T$; the dashed lines give the K41 result for these exponent ratios; in Table
\ref{tab:ESSZetas} we give the numerical values of these exponents ratios
(along with error bars, which we determine by a local-slope analysis). From
Figs.~\ref{fig:PiZeta} (c) and (d), we observe that $\zeta^{\rm
n}_p/\zeta^3_p$, $\zeta^{\rm s}_p/\zeta^{\rm s}_3 > p/3$ for $p = 1$ to $2$,
and $\zeta^{\rm n}_p/\zeta^3_p$, $\zeta^{\rm s}_p/ \zeta^{\rm s}_3 <  p/3$ for
$p = 4$ to $6$; these are clear signatures of intermittency in superfluid 
turbulence. Furthermore, we observe that the values of
the ratios $\zeta^{\rm n}_p/\zeta^3_p$ and $\zeta^{\rm s}_p/\zeta^{\rm s}_3 $ differ
most from their K41 values in the temperature range $T = 1.7$ to $1.9$. We
can characterize the intermittency by the exponents (see, e.g.,
Ref.~\cite{dhar1997some}) $\mu_p^{\rm n} = p/3 - \zeta_6^{\rm n}$; $\mu_p^{\rm
s} = p/3 - \zeta_6^{\rm s}$, for $p = 5$ and $6$,  which measure the deviation
of the $5^{th}-$ and $6^{th}-$order exponents from their K41 values. In
Figs.~\ref{fig:PiZeta} (e) and(f) we plot $\mu_p^{\rm n}$, and $\mu_p^{\rm s}$,
respectively, for $p = 5$ (red lines) and $p = 6$ (blue lines). From
Figs.~\ref{fig:PiZeta} (e), and (f) we observe that these deviations, and hence
the intermittency, are highest in the temperature range $T = 1.70$ to $ T =
1.90$. Figures~\ref{fig:PiZeta} (g), (h), and (i) are similar to Figs.\ref{fig:PiZeta} (a), (c), and (e), but for the runs with temperature dependent viscosities (Table~\ref{tab:parameters}). Intermittency in superfluid turbulence has also been studied in
Refs.~\cite{salort2011investigation,
rusaouen2017intermittency,procacciaintermittencyshellmodel,
shukla2016multiscaling,biferale2018turbulent,varga2018intermittency}
experimentally and numerically, by shell-model and DNS studies of 3D HVBK
turbulence. As in classical-fluid turbulence, we still lack an
\textit{ab-initio} theory of such intermittency.

\begin{table*} [t]
\begin{tabular}{c  c  c  c  c  c  c  c  c  c  c  c  }
\hline
Run & ${\zeta_{1}^{\rm n}}/{\zeta_{3}^{\rm n}} $ & $\zeta_{2}^{\rm n}/{\zeta_{3}^{\rm n}} $ & $\zeta_{4}^{\rm n}/{\zeta_{3}^{\rm n}} $ & $\zeta_{5}^{\rm n}/{\zeta_{3}^{\rm n}}$ & $\zeta_{6}^{\rm n}/{\zeta_{3}^{\rm n}} $ & ${\zeta_{1}^{\rm s}}/{\zeta_{3}^{\rm s}} $ & $\zeta_{2}^{\rm s}/{\zeta_{3}^{\rm s}} $ & $\zeta_{4}^{\rm s}/{\zeta_{3}^{\rm s}} $ & $\zeta_{5}^{\rm s}/{\zeta_{3}^{\rm s}}$ & $\zeta_{6}^{\rm s}/{\zeta_{3}^{\rm s}} $ \\
\hline \hline

\bf R1 & $0.36 \pm 0.00$ & $0.70 \pm 0.00$ & $1.27 \pm 0.00$ & $1.52 \pm 0.00$
& $1.75 \pm 0.01$ & $0.36 \pm 0.00$ & $0.70 \pm 0.00$ & $1.27 \pm 0.00$
& $1.52 \pm 0.01$ & $1.75 \pm 0.01$ \\
\hline

\bf R2 & $0.37 \pm 0.00$ & $0.70 \pm 0.00$ & $1.27 \pm 0.00$ & $1.52 \pm 0.01$
& $1.75 \pm 0.02$ & $0.37 \pm 0.00$ & $0.70 \pm 0.00$ & $1.27 \pm 0.00$
& $1.52 \pm 0.02$ & $1.74 \pm 0.03$ \\
\hline

\bf R3 & $0.37 \pm 0.00$ & $0.70 \pm 0.00$ & $1.27 \pm 0.01$ & $1.52 \pm 0.02$
& $1.75 \pm 0.04$ & $0.37 \pm 0.01$ & $0.70 \pm 0.01$ & $1.27 \pm 0.02$
& $1.51 \pm 0.04$ & $1.73 \pm 0.07$ \\
\hline

\bf R4 & $0.37 \pm 0.00$ & $0.71 \pm 0.00$ & $1.26 \pm 0.01$ & $1.48 \pm 0.01$
& $1.68 \pm 0.02$ & $0.37 \pm 0.00$ & $0.70 \pm 0.00$ & $1.26 \pm 0.01$
& $1.49 \pm 0.01$ & $1.69 \pm 0.02$ \\
\hline

\bf R5 & $0.37 \pm 0.00$ & $0.70 \pm 0.00$ & $1.27 \pm 0.01$ & $1.51 \pm 0.01$
& $1.73 \pm 0.02$ & $0.37 \pm 0.00$ & $0.70 \pm 0.01$ & $1.27 \pm 0.00$
& $1.52 \pm 0.01$ & $1.74 \pm 0.02$ \\
\hline

\bf R6 & $0.37 \pm 0.01$ & $0.70 \pm 0.01$ & $1.26 \pm 0.01$ & $1.50 \pm 0.03$
& $1.70 \pm 0.05$ & $0.37 \pm 0.00$ & $0.70 \pm 0.01$ & $1.27 \pm 0.01$
& $1.50 \pm 0.03$ & $1.71 \pm 0.05$ \\
\hline

\bf R7 & $0.36 \pm 0.00$ & $0.69 \pm 0.00$ & $1.28 \pm 0.00$ & $1.54 \pm 0.01$
& $1.77 \pm 0.01$ & $0.36 \pm 0.00$ & $0.70 \pm 0.00$ & $1.27 \pm 0.01$
& $1.52 \pm 0.02$ & $1.74 \pm 0.04$ \\
\hline

\bf R8 & $0.36 \pm 0.00$ & $0.70 \pm 0.00$ & $1.28 \pm 0.00$ & $1.52 \pm 0.01$
& $1.75 \pm 0.02$ & $0.36 \pm 0.00$ & $0.70 \pm 0.00$ & $1.28 \pm 0.00$
& $1.53 \pm 0.01$ & $1.75 \pm 0.02$ \\
\hline
\bf R9 & $0.33 \pm 0.04$ & $0.67 \pm 0.03$ & $1.31 \pm 0.05$ & $1.58 \pm 0.09$
& $1.81 \pm 0.14$ & $0.35 \pm 0.03$ & $0.68 \pm 0.02$ & $1.29 \pm 0.03$
& $1.56 \pm 0.08$ & $1.79 \pm 0.13$ \\
\hline
\bf R10 & $0.31 \pm 0.03$ & $0.66 \pm 0.02$ & $1.32 \pm 0.03$ & $1.59 \pm 0.07$
& $1.82 \pm 0.10$ & $0.35 \pm 0.01$ & $0.68 \pm 0.01$ & $1.29 \pm 0.01$
& $1.56 \pm 0.03$ & $1.82 \pm 0.05$ \\
\hline
\bf R11 & $0.36 \pm 0.02$ & $0.69 \pm 0.02$ & $1.27 \pm 0.02$ & $1.52 \pm 0.05$
& $1.73 \pm 0.08$ & $0.35 \pm 0.02$ & $0.69 \pm 0.01$ & $1.28 \pm 0.02$
& $1.53 \pm 0.04$ & $1.76 \pm 0.06$ \\
\hline
\bf R12 & $0.37 \pm 0.02$ & $0.70 \pm 0.02$ & $1.26 \pm 0.04$ & $1.49 \pm 0.11$
& $1.69 \pm 0.2$ & $0.37 \pm 0.01$ & $0.69 \pm 0.01$ & $1.28 \pm 0.01$
& $1.53 \pm 0.02$ & $1.77 \pm 0.04$ \\
\hline
\bf R13 & $0.37 \pm 0.02$ & $0.70 \pm 0.02$ & $1.26 \pm 0.05$ & $1.49 \pm 0.12$
& $1.70 \pm 0.21$ & $0.37 \pm 0.02$ & $0.70 \pm 0.02$ & $1.26 \pm 0.05$
& $1.49 \pm 0.12$ & $1.70 \pm 0.21$ \\
\hline
\bf R14 & $0.36 \pm 0.06$ & $0.69 \pm 0.05$ & $1.28 \pm 0.07$ & $1.52 \pm 0.16$
& $1.75 \pm 0.25$ & $0.36 \pm 0.02$ & $0.69 \pm 0.02$ & $1.27 \pm 0.03$
& $1.51 \pm 0.07$ & $1.72 \pm 0.12$ \\
\hline

\end{tabular} 
\caption{The numerical values of the exponent ratios $\zeta_p^{\rm n}/\zeta_3^{\rm n}$ and $\zeta_p^{\rm s}/\zeta_3^{\rm s}$, from all our DNSs $R1-R14$ and
for $1 \leq p \leq 6$, along  with error bars, which we determine by a 
local-slope analysis. To determine  these exponent ratios we use the 
extended-self-similarity (ESS) method (see text).}
\label{tab:ESSZetas}

\end{table*}

\section{ CONCLUSIONS} \label{sec:Conclusions}

We have used the generating-functional approach to derive the von K\'arm\'an-Howarth 
relations [Eqs.~(\ref{eq:KH1})-(\ref{eq:KH4})] for the 3D HVBK model of superfluid turbulence; 
and we have shown that the simple von K\'arm\'an-Howarth relation,
for classical-fluid turbulence, is replaced by four relations here. In particular,
we have included the effects of the mutual-friction term (if this term is neglected,
our general results reduce to those in Ref.~\cite{biferale2018turbulent}).
Furthermore, we have obtained power-law behaviors for the PDFs $P(\gamma)$ 
and $P(\theta)$ from our DNS results; we have then shown how these power laws can be 
understood analytically, if we make reasonable decoupling approximations for 
certain joint PDFs. The exponents of $P(\gamma)$ for the 2D HVBK case, which  
have been calculated numerically in Ref.~\cite{shukla2015homogeneous},
are in good agreement with our analytical predictions. These power-law exponents are 
universal in the sense that they are independent of the mutual-friction coefficients 
$B$ and $B'$ and the temperature $T$; it should be possible to measure them in 
experiments, such as those conducted in Refs.~\cite{salort2011investigation,rusaouen2017intermittency,varga2018intermittency} for superfluid $^4$He.

From our DNSs we have obtained energy, energy-flux, and
mutual-friction-function spectra.  the longitudinal-structure-function
exponents for the normal fluid and the superfluid, as a function of the
temperature $T$. We have calculated the ratios of structure-function exponents
for the normal fluid and the superfluid, via the ESS method, as a function of
$T$, by using the experimentally determined mutual-friction coefficients for
superfluid Helium $^4$He~\cite{donnelly1998omfdata}. We have shown that there
is an enhancement of intermittency for the normal fluid and the superfluid in
the range $1.7 \leq T \leq 1.90$; our results should be applicable to, and
verifiable in, experiments like those of
Refs.~\cite{salort2011investigation,rusaouen2017intermittency,varga2018intermittency};
they are also similar to the intermittency results in the DNSs of
Ref.~\cite{biferale2018turbulent}.


{\bf Acknowledgments} We thank A. Bhatnagar and P.E. Roche for fruitful
discussions, CSIR, UGC, DST, SERB, and the National Supercomputing Mission (India) for financial support, and SERC (IISc)
for providing computational resources. A.B. thanks the Alexander von Humboldt
Stiftung, Germany for partial financial support through the Research Group
Linkage Programme (2016).

\onecolumngrid
\noindent\rule{18cm}{0.4pt} \\

\section{Appendix}

We give below some details of our calculations for the structure-function hierarchy.

The pressure contribution, from the normal fluid, is: 
\begin{eqnarray}
\eta_{{\rm n}3}I_{\rm p} = \eta_{{\rm n}3}\bigg \langle \bigg[{\boldsymbol 
\lambda}_{\rm n} \cdot \frac{1}{\rho_{\rm n}} \nabla (\Delta p_{\rm n}) \bigg ]
Z_{\rm n} \bigg \rangle; 
\end{eqnarray}

\begin{equation}
\eta_{{\rm n}3}I_{\rm p} = \bigg \langle \bigg [\eta_{{\rm n}2}\eta_{{\rm n}3}
\frac{1}{\rho_{\rm n}} \big(\nabla (\Delta p_{\rm n})\big)_{\parallel} + 
\eta_{{\rm n}3}^2 \frac{1}{\rho_{\rm n}} \big(\nabla (\Delta p_{\rm n})\big)_{\perp} \bigg ] Z_{\rm n} \bigg \rangle .
\label{eq:Ip_n1}
\end{equation}
If we take the derivative $\partial_{\eta_{{\rm n}2}}^2 \partial_{\eta_{{\rm n}3}}$ of Eq.~(\ref{eq:Ip_n1}) and the limits $\eta_{{\rm n}2},\eta_{{\rm n}3}
\rightarrow 0$, we get
\begin{equation}
\lim_{\eta_{{\rm n}2},\eta_{{\rm n}3}\rightarrow 0}\bigg(\partial_{\eta_{{\rm n}2}}^2 \partial_{\eta_{{\rm n}3}}\big (\eta_{{\rm n}3}I_p \big) \bigg) =
\frac{1}{\rho_{\rm n}}\bigg \langle \Delta u_{{\rm n}_{\parallel}}\big(\nabla
(\Delta p_{\rm n}) \big)_{\parallel} \bigg \rangle .
\end{equation}

By applying the derivative $\partial_{\eta_{{\rm n}3}}^3 $ on Eq.~(\ref{eq:Ip_n1}), 
and after taking the limits $\eta_{{\rm n}2},\eta_{{\rm n}3} \rightarrow 0$, we get
\begin{equation}
\lim_{\eta_{{\rm n}3},\eta_{{\rm n}2}\rightarrow 0}\bigg(\partial_{\eta_{{\rm n}2}}^3 \big (\eta_{{\rm n}3}I_{\rm p} \big) \bigg) = \frac{1}{\rho_{\rm n}}\bigg \langle \Delta u_{{\rm n}_{\perp}} \big(\nabla (\Delta p_{\rm n})\big)_{\perp} \bigg \rangle .
\end{equation}

The total pressure contribution to the third-order structure function for the normal
fluid is 
\begin{equation}
\frac{1}{\rho_{\rm n}}\bigg \langle \Delta u_{{\rm n}_{\parallel}} \big(\nabla
(\Delta p_{\rm n})\big)_{\parallel} + \Delta u_{{\rm n}_{\perp}} \big(\nabla (\Delta p_{\rm n}) \big)_{\perp} \bigg \rangle = \frac{1}{\rho_{\rm n}}
\bigg \langle  \Delta u_{{\rm n}i} \nabla_{i} \Delta p_{\rm n} \bigg \rangle ;
\end{equation}

\begin{equation}
\bigg \langle \Delta u_{{\rm n}_{\parallel}} \big(\nabla(\Delta p_{\rm n})\big)_{\parallel} + \Delta u_{{\rm n}_{\perp}} \big(\nabla (\Delta p_{\rm n}) 
\big)_{\perp} \bigg \rangle = \bigg \langle  u_{{\rm n}i}({\bf x}1) \nabla_{i} 
p_{\rm n} ({\bf x}1) - u_{{\rm n}i}({\bf x}2) 
\nabla_{i} p_{\rm n} ({\bf x}1) - u_{{\rm n}i}({\bf x}1) \nabla_{i} p_{\rm n} 
({\bf x}2) + u_{{\rm n}i}({\bf x}2) \nabla_{i} p_{\rm n} ({\bf x}2) \bigg
\rangle .
\label{eq:Ip_n2}
\end{equation}

In the RHSs of the above equations, we have contributions from the followng two
types of terms: (1) terms at the same point, and (2) terms at two different
points. By using the homogeneity condition, we write $ \big \langle  u_{{\rm
n}i}({\bf x}1) \nabla_{i} p_{\rm n} ({\bf x}1) \big \rangle = \nabla_{i} \big
\langle  u_{{\rm n}i}({\bf x}1) p_{\rm n} ({\bf x}1) \big \rangle .$ From  
the condition of (statistical) homogeneity, we get $ \nabla_{i} \big \langle
u_{{\rm n}i}({\bf x}1) p_{\rm n} ({\bf x}1) \big \rangle = 0.$ Similarly, we
get $\nabla_{i} \big \langle u_{{\rm n}i}({\bf x}2) p_{\rm n} ({\bf x}2) \big
\rangle = 0.$ By using the incompressibility condition, we write $\big \langle
u_{{\rm n}i} ({\bf x}1) \nabla_{i} p_{\rm n} ({\bf x}2) \big \rangle =
\nabla_i({\bf x}1) \big \langle u_{{\rm n}i} ({\bf x}1) p_{\rm n} ({\bf x}2)
\big \rangle = 0.$ We define ${\bf r} = {\bf x}1 - {\bf x}2 $ and this gives us
$\nabla_i ({\bf r}) = \nabla_i({x}1)$. If we  apply the homogeneity condition
and consider that $\big \langle u_{{\rm n}i} ({\bf x}1) p_{\rm n} ({\bf x}2)
\big \rangle = A(r) \frac{r_i}{r} $, then the physical solution of $\nabla_{i}
({\bf x}_1) \big( A(r) \frac{r_i}{r}\big) = 0 $ is $A(r) = 0$. Thus, $\big
\langle u_{ni}({\bf x}1) p({\bf x}2)\big \rangle = 0$; similarly, we can get
$\big \langle u_{{\rm n}i}({\bf x}2) p({\bf x}1)\big \rangle = 0$. Now
Eq.~(\ref{eq:Ip_n2}) becomes

\begin{equation}
\bigg \langle \Delta u_{{\rm n}_{\parallel}}\big(\nabla(\Delta p_{\rm n})\big)_{\parallel}\bigg \rangle + \bigg \langle \Delta u_{{\rm n}_{\perp}} \big(\nabla 
(\Delta p_{\rm n}) \big)_{\perp} \bigg \rangle = 0 .
\end{equation}

The contribution from the perpendicular component in the above equation can be
written as $ \big \langle \Delta u_{{\rm n}_{\perp}} \big(\nabla (\Delta p_{\rm
n}) \big)_{\perp} \big \rangle = \big \langle \Delta u_{{\rm
n}_{\theta}}\frac{1}{r} \frac{\partial}{\partial \theta} \Delta p_{\rm n} \big
\rangle.$ The term $ \big ( \Delta u_{{\rm n}_{\theta}} \frac{1}{r}
\frac{\partial} {\partial \theta} \Delta p_{\rm n} \big) $ changes its sign
under the replacement $\theta \rightarrow -\theta$, hence $ \big \langle \Delta u_{{\rm
n}_{\theta}} \frac{1}{r} \frac{\partial}{\partial \theta} \Delta p_{\rm n} \big
\rangle = 0.$ Furthermore, it implies that $ \big \langle \Delta u_{{\rm
n}_{\parallel}} \big(\nabla (\Delta p_{\rm n}) \big)_{\parallel} \big \rangle =
\big \langle \Delta u_{{\rm n}_{\perp}} \big(\nabla (\Delta p_{\rm n})
\big)_{\perp} \big \rangle= 0.$ Similarly, we can show that pressure
contribution from the superfluid components is also zero, i.e.,  $ \big \langle
\Delta u_{{\rm s}_{\parallel}} \big(\nabla (\Delta p_{\rm s})
\big)_{\parallel} \big \rangle = \big \langle \Delta u_{{\rm s}_{\perp}}
\big(\nabla (\Delta p_{\rm s}) \big)_{\perp} \big \rangle= 0.$ Thus, the
pressure term does not contribute to the third-order structure functions.

The dissipation term for normal fluid is given as
\begin{eqnarray}
D_{\rm n} = \Big \langle \nu_{\rm n}\Big[{\boldsymbol \lambda}_{{\rm n}1} 
\cdot \nabla^2_{{\bf x}_1} {\bf u}_{\rm n}{({\bf x}_1)}+ {\boldsymbol 
\lambda}_{{\rm n}2} \cdot\nabla^2_{{\bf x}_2} {\bf u}_{\rm n}{({\bf x}_2)}\Big]
Z_{\rm n} \Big \rangle .
\end{eqnarray}

For convenience, we consider that ${\boldsymbol \lambda}_{{\rm n}1} = 
-{\boldsymbol \lambda}_{{\rm n}2}$  and ${\boldsymbol \lambda}_{{\rm s}1} =
-{\boldsymbol \lambda}_{{\rm s}2}$; and for notational simplicity we consider
${\boldsymbol \lambda}_{{\rm n}1}={\boldsymbol \lambda}_{\rm n}$ and 
${\boldsymbol \lambda}_{{\rm s}1} =  {\boldsymbol \lambda}_{\rm s}$. In terms
of ${\boldsymbol \lambda}_{\rm n}$ and ${\boldsymbol \lambda}_{\rm s}$  the
dissipation terms are
\begin{eqnarray}
D_{\rm n} =\Big\langle \nu_{\rm n} \Big[{\boldsymbol \lambda}_{\rm n}\cdot
\nabla^2_{{\bf x}_1} {\bf u}_{\rm n}{({\bf x}_1)}-{\boldsymbol \lambda}_{\rm n}
\cdot\nabla^2_{{\bf x}_2} {\bf u}_{\rm n}{({\bf x}_2)}\Big] Z_{\rm n} \Big
\rangle ;
\end{eqnarray}

\begin{eqnarray}
D_{\rm n}=\nu_{\rm n}\Big\langle \Big[\eta_{{\rm n}2}\nabla^2_{{\bf x}_1}{\bf u}_{{\rm n}_\parallel} {({\bf x}_1)}-\eta_{{\rm n}2} \nabla^2_{{\bf x}_2} 
{\bf u}_{{\rm n}_\parallel}{({\bf x}_2)}+ \eta_{{\rm n}3}\nabla^2_{{\bf x}_1} 
{\bf u}_{{\rm n}_\perp} {({\bf x}_1)}-\eta_{{\rm n}3} \nabla^2_{{\bf x}_2} 
{\bf u}_{{\rm n}_\perp}{({\bf x}_2)}\Big] Z_{\rm n} \Big \rangle .
\end{eqnarray}

We note that 
\begin{align}
& \Big \langle \nu_{\rm n} \Big (\nabla^2_{{\bf x}_1}+\nabla^2_{{\bf x}_2}\Big)
Z_{\rm n} \Big \rangle = \nu_{\rm n} \langle  \Big [\eta_{{\rm n}2}\nabla^2_{
{\bf x}_1} {\bf u}_{{\rm n}_\parallel} {({\bf x}_1)}-\eta_{{\rm n}2} 
\nabla^2_{{\bf x}_2} {\bf u}_{{\rm n}_\parallel}{({\bf x}_2)}+ \eta_{{\rm n}3}
\nabla^2_{{\bf x}_1} {\bf u}_{{\rm n}_\perp} {({\bf x}_1)} -\eta_{{\rm n}3} 
\nabla^2_{{\bf x}_2} {\bf u}_{{\rm n}_\perp}{({\bf x}_2)}\Big] Z_{\rm n}  \Big
\rangle   + \nonumber \\ & + \nu_{\rm n} \Big \langle Z_{\rm s} \Big 
[\eta_{{\rm n}2}\nabla_{{\bf x}_1} {\bf u}_{{\rm n}_\parallel} {({\bf x}_1)}+
\eta_{{\rm n}3}\nabla_{{\bf x}_1} {\bf u}_{{\rm n}_\perp} {({\bf x}_1)}\Big]^2
Z_{\rm n} \Big \rangle + \nu_{\rm n} \Big \langle Z_{\rm s} \Big 
[\eta_{{\rm n}2}\nabla_{{\bf x}_2}{\bf u}_{{\rm n}_\parallel} {({\bf x}_2)}+
 \eta_{{\rm n}3}\nabla_{{\bf x}_2} {\bf u}_{{\rm n}_\perp} {({\bf x}_2)}\Big]^2 Z_{\rm n} \Big \rangle ;
\end{align}
by substituting the value of $D_{\rm n}$ in this equation, we get 

\begin{align}
& \Big \langle  \nu_{\rm n} \Big(\nabla^2_{{\bf x}_1}+\nabla^2_{{\bf x}_2}\Big)
Z_{\rm n} \Big \rangle  =  D_{\rm n} + \nu_{\rm n} \Big \langle \Big
[\eta_{{\rm n}2}^2 \Big (\nabla_{{\bf x}_1} {\bf u}_{{\rm n}_\parallel}
{({\bf x}_1)}\Big)^2 + \eta_{{\rm n}3}^2 \Big (\nabla_{{\bf x}_1} 
{\bf u}_{{\rm n}_\perp} {({\bf x}_1)}\Big)^2 + 2 \eta_{{\rm n}2} \eta_{{\rm n}3}
\Big(\nabla_{{\bf x}_1} {\bf u}_{{\rm n}_\parallel} {({\bf x}_1)}\Big)\Big
(\nabla_{{\bf x}_1} {\bf u}_{{\rm n}_\perp}{({\bf x}_1)}\Big)
\Big]Z_{\rm n} \Big \rangle \nonumber \\ & +  \nu_{\rm n} \Big \langle  \Big
[\eta_{{\rm n}2}^2  \Big (\nabla_{{\bf x}_2} {\bf u}_{{\rm n}_\parallel}
{({\bf x}_2)}\Big)^2 +\eta_{{\rm n}3}^2 \Big(\nabla_{{\bf x}_2}
{\bf u}_{{\rm n}_\perp} {({\bf x}_2)}\Big)^2 + 2 \eta_{{\rm n}2} \eta_{{\rm n}3}\Big(\nabla_{{\bf x}_2} {\bf u}_{{\rm n}_\parallel} {({\bf x}_2)}\Big) 
\Big(\nabla_{{\bf x}_2} {\bf u}_{{\rm n}_\perp}{({\bf x}_2)}\Big)
\Big ]Z_{\rm n} \Big \rangle .
\end{align}
On using $\nu_{\rm n}\Big(\nabla_{{\bf a}}{\bf u}_{{\rm n}_\parallel}
{({\bf a})}\Big)^2 = \epsilon_{{\rm n}_\parallel}({\bf a})$ and  $\nu_{\rm n}
\Big(\nabla_{{\bf a}}{\bf u}_{{\rm n}_\perp} {({\bf a})}\Big)^2 = 
\epsilon_{{\rm n}_\perp}({\bf a})$, where ${\bf a}$ stands for ${\bf x}_1$ or
${\bf x}_2$, we get the following:

\begin{align}
&\Big \langle \nu_{\rm n} \Big(\nabla^2_{{\bf x}_1}+\nabla^2_{{\bf x}_2}\Big)
Z_{\rm n} \Big \rangle = D_{\rm n} + \Big \langle \Big [\eta_{{\rm n}2}^2
\Big(\epsilon_{{\rm n}_\parallel}({\bf x}_1)+\epsilon_{{\rm n}_\parallel}
({\bf x}_2)\Big)+ \eta_{{\rm n}3}^2\Big(\epsilon_{{\rm n}_\perp}({\bf x}_1) +
\epsilon_{{\rm n}_\perp}({\bf x}_2)\Big)\Big] Z_{\rm n}  \Big \rangle 
\nonumber \\ & + 2 \Big \langle \eta_{{\rm n}2}\eta_{{\rm n}3}\Big 
[\Big(\epsilon_{{\rm n}_\parallel} {({\bf x}_1)} \epsilon_{{\rm n}_\perp}
{({\bf x}_1)}\Big)^{\frac{1}{2}} \Big(\epsilon_{{\rm n}_\parallel}{({\bf x}_2)}
\epsilon_{{\rm n}_\perp} {({\bf x}_2)}\Big)^{\frac{1}{2}}\Big] Z_{\rm n} 
\Big \rangle .
\end{align}

If we take the limit $\nu_{\rm n} \rightarrow 0$ and set $\epsilon_{{\rm
n}_\parallel}({\bf x}_1)+\epsilon_{{\rm n}_\parallel} ({\bf x}_2)=
\epsilon_{{\rm n}_\parallel}$ and $ \epsilon_{{\rm n}_\perp} ({\bf
x}_1)+\epsilon_{{\rm n}_\perp} ({\bf x}_2)=\epsilon_{{\rm n}_\perp}$, in the
above equation, we get

\begin{align}
0 = & D_{\rm n} +\Big \langle\Big [\eta_{{\rm n}2}^2\epsilon_{{\rm n}_\parallel}
+\eta_{{\rm n}3}^2 \epsilon_{{\rm n}_\perp} \Big ]Z_{\rm n} \Big \rangle + 
2 \Big \langle \eta_{{\rm n}2}\eta_{{\rm n}3}\Big[\Big(\epsilon_{{\rm n}_\parallel}{({\bf x}_1)} \epsilon_{{\rm n}_\perp}{({\bf x}_1)}\Big)^{\frac{1}{2}}
+ \Big(\epsilon_{{\rm n}_\parallel}{({\bf x}_2)}\epsilon_{{\rm n}_\perp}
{({\bf x}_2)}\Big)^{\frac{1}{2}}\Big] Z_{\rm n} \Big \rangle ;
\end{align}
or
\begin{align}
-D_{\rm n}= & \Big \langle\Big [\eta_{{\rm n}2}^2\epsilon_{{\rm n}_\parallel}
+\eta_{{\rm n}3}^2 \epsilon_{{\rm n}_\perp} ]Z_{\rm n}  \Big \rangle +
2 \Big \langle \eta_{{\rm n}2}\eta_{{\rm n}3}\Big[\Big(\epsilon_{{\rm n}_\parallel} {({\bf x}_1)} \epsilon_{{\rm n}_\perp}{({\bf x}_1)}\Big)^{\frac{1}{2}}
+ \Big(\epsilon_{{\rm n}_\parallel}{({\bf x}_2)}\epsilon_{{\rm n}_\perp}
{({\bf x}_2)}\Big)^{\frac{1}{2}}\Big] Z_{\rm n}  \Big \rangle .
\end{align}
Similarly, the dissipation term from the superfluid part is 

\begin{align}
-D_{\rm s} = & \Big \langle\Big [\eta_{{\rm s}2}^2\epsilon_{{\rm s}_\parallel}
+\eta_{{\rm s}3}^2  \epsilon_{{\rm s}_\perp} ]Z_{\rm s} \Big \rangle + 
2 \Big \langle \eta_{{\rm s}2}\eta_{{\rm s}3}\Big[\Big(\epsilon_{{\rm s}_\parallel} {({\bf x}_1)} \epsilon_{{\rm s}_\perp}{({\bf x}_1)}\Big)^{\frac{1}{2}}
+ \Big(\epsilon_{{\rm s}_\parallel}{({\bf x}_2)}\epsilon_{{\rm s}_\perp}
{({\bf x}_2)}\Big)^{\frac{1}{2}}\Big] Z_{\rm s}  \Big \rangle .
\end{align}

\twocolumngrid

\end{document}